\documentclass[review]{elsarticle}

\usepackage{lineno,hyperref}
\modulolinenumbers[5]
\usepackage[nodots]{numcompress}
\biboptions{numbers,sort&compress}

\usepackage{amsmath}
\usepackage{amsfonts}
\usepackage{amssymb}
\usepackage{color, soul}

\usepackage{multirow}
\journal{Optics and Lasers in Engineering}

\begin{document}
\begin{frontmatter}
\title{Reconstructing Stokes parameters from non-uniform division-of-focal-plane modulation}

\author[mymainaddress]{Zhaoxiang Jiang}

\author[mymainaddress]{Qingchuan Zhang\corref{mycorrespondingauthor1}}
\cortext[mycorrespondingauthor1]{Corresponding author}
\ead{zhangqc@ustc.edu.cn}

\author[mymainaddress]{Shangquan Wu\corref{mycorrespondingauthor}}
\cortext[mycorrespondingauthor]{Corresponding author}
\ead{wushq@ustc.edu.cn}

\author[mymainaddress]{Tan Xu}

\author[mymainaddress]{Yong Su}

\author[mymainaddress]{Chuanbiao Bai}

\address[mymainaddress]{CAS Key Laboratory of Mechanical Behavior and Design of Materials, Department of Modern Mechanics, University of Science and Technology of China, Hefei 230027, China}

\begin{abstract}
Division-of-focal-plane modulation is a powerful technique for real-time polarization imaging. This technique, however, suffers from the non-uniformity of the performance of linear polarization filters and photodetectors. We study the Stokes parameters reconstruction from the division-of-focal-plane modulation in the presence of the non-uniformity. Two reconstruction methods, named as ordinary least-squares and smoothing regularization methods, are proposed. The performance of the proposed methods are evaluated through Fourier analysis, numerical simulations, and experiments. The results indicate that the proposed methods can effectively mitigate the reconstruction errors and artifacts caused by the non-uniformity, and therefore may further facilitate the practical application of the division-of-focal-plane technique.
\end{abstract}


\end{frontmatter}


\section{Introduction}\label{sec1}
\begin{figure}[htb!]
	\centering\includegraphics[width=7cm]{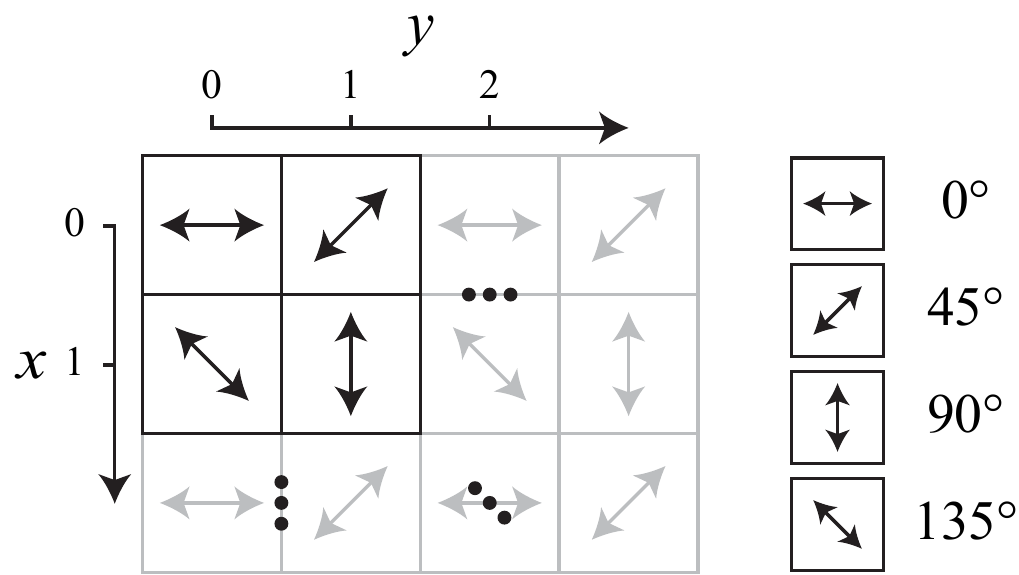}
	\caption{Polarization filter array arrangement composed by 2$\times$2 periodically patterned 0$^{\circ}$, 45$^{\circ}$, 90$^{\circ}$, and 135$^{\circ}$ linear polarization filters.}
	\label{fig1} 
\end{figure}

Polarization imaging aims to measure the polarization information described by Stokes parameters and their derivatives such as degree of linear polarization (DoLP) and angle of polarization (AoP). It has a wide range of applications in various fields such as remote sensing~\cite{RN1742}, biomedical diagnosis~\cite{RN1741,RN1745}, and interferometry~\cite{RN93,RN1546,RN1545,RN1748,RN1747,RN1749}. The modulation technique in polarization imaging can be classified into four categories: division-of-time (DoT)~\cite{RN1736}, division-of-amplitude~\cite{RN1737,RN1746},  division-of-aperture~\cite{RN1735}, and division-of-focal-plane (DoFP)~\cite{RN833,RN1725,RN1726}. The DoFP technique achieves the polarization modulation by integrating a polarization filter array in front of a photodetector array. The most commonly used polarization filter array arrangement is shown in Fig.~\ref{fig1}, which is composed by 2$\times$2 periodically patterned 0$^{\circ}$, 45$^{\circ}$, 90$^{\circ}$, and 135$^{\circ}$ linear polarization filters. The DoFP polarimeters have advantages of compact structure and high temporal resolution, and therefore are especially suitable for real-time polarization imaging.

A fundamental problem in the DoFP technique is the Stokes parameters reconstruction. Since each pixel can only capture the polarization information in one orientation, the measurement of the Stokes parameters is incomplete. Mathematically, reconstructing the Stokes parameters from the DoFP polarization modulation is an ill-posed inverse problem. Many methods have been proposed to reconstruct the Stokes parameters~\cite{RN1724,RN836,RN1652,RN1593,RN1047,RN1665,RN1051}. Most of these methods are spatial domain interpolation-based methods~\cite{RN1724,RN836,RN1652,RN1593} and frequency domain filtering-based methods~\cite{RN1047,RN1665,RN1051}. In the interpolation-based methods, the DoFP image is split into 0$^{\circ}$, 45$^{\circ}$, 90$^{\circ}$, and 135$^{\circ}$ polarization images. After interpolating the missing pixel values in these polarization images, the interpolation-based methods determine the Stokes parameters through the ordinary least-squares criterion. The interpolation algorithms are usually convolutional, including nearest-neighbor, bilinear, bicubic, and natural bicubic spline interpolation algorithms~\cite{RN1724,RN836}. Some edge-preserved interpolation algorithms have also been proposed recently~\cite{RN1652}. Based on the characteristics of the spectrum of the DoFP image, the filtering-based methods use the filter transfer functions constructed by window functions to reconstruct the Stokes parameters. The window functions used in previous studies including Hamming~\cite{RN1047}, Gaussian~\cite{RN1665}, and Planck-taper~\cite{RN1051} window functions. However, the interpolation-based and filtering-based methods are only suitable for the theoretical case that the Stokes parameters are periodically modulated. In practice, the manufacturing imperfects of the DoFP polarimeters cause the non-uniformity of the performance of the linear polarization filters and photodetectors. The non-uniformity is characterized as the differences of the major and minor principal transmittances of the linear polarization filters, the differences of the gains and dark offsets of the photodetectors, and the deviations between the actual and designed orientations of the linear polarization filters~\cite{RN1502,RN834,RN1720}. The non-uniformity destroys the periodicity of the polarization modulation. Consequently, the interpolation-based and filtering-based methods will fail in practical applications since these methods are unable to tackle the reconstruction errors and artifacts caused by the non-uniformity.

In this paper, we study the Stokes parameters reconstruction from the DoFP modulation in the presence of the non-uniformity. We propose two reconstruction methods that can tackle the reconstruction errors and artifacts caused by the non-uniformity. The proposed methods are inspired by the classical Lucas-Kanade method~\cite{RN1727} and Horn-Schunck method~\cite{RN1728} in optical flow estimation. One is the ordinary least-squares method (OLSM), which reconstructs the Stokes parameters under the local constant assumption that the Stokes parameters are constant functions in 2$\times$2 subsets. The basic idea of this method has been reported in Ref.~\cite{david2008unpolarized, RN1643} and our previous study~\cite{RN1637}. Here, we further add a four-subset averaging strategy, present in-depth theoretical analyses, and explain the relationship between the OLSM and interpolation-based methods. The other is the smoothing regularization method (SRM), which reconstructs the Stokes parameters under the global smoothing assumption that the Stokes parameters are spatially smooth. This method has more similarities to the filtering-based methods.

This paper is organized as follows: in Section~\ref{sec2}, a linear pixel model is introduced to characterize the non-uniformity; in Section~\ref{sec3}, the OLSM and SRM are presented to reconstruct the Stokes parameters; in Section~\ref{sec4}, Fourier analysis and numerical simulations are used to evaluate the reconstruction errors of the OLSM and SRM; in Section~\ref{sec5}, the choice of the regularization parameters in the SRM is discussed; in Section~\ref{sec6}, the performance of the OLSM, SRM, and interpolation-based and filtering-based methods is evaluated and compared through two experiments; in Section~\ref{sec7}, the performance of the OLSM and SRM is summarized.
\section{Linear pixel model}\label{sec2}
After integrating with the array composed by the linear polarization filters, the photodetector array can be regarded as being sensitive to the first three Stokes parameters. Assuming the performance of the linear polarization filters and photodetectors is linear, for each pixel, we have~\cite{RN834}
\begin{equation}\label{eq1}
i(x,y)=m_0(x,y)s_0(x,y)+m_1(x,y)s_1(x,y)+m_2(x,y)s_2(x,y)+d(x,y).
\end{equation}
Here, $x$ and $y$ are the horizontal and vertical pixel coordinates, respectively, $i$ represents the intensity of the DoFP image, $s_0$, $s_1$, and $s_2$ represent the first three Stokes parameters, $m_0$, $m_1$, and $m_2$ are the modulation parameters of $s_0$, $s_1$, and $s_2$, respectively, and $d$ represents the dark offset of the photodetector. Specifically, $m_0$, $m_1$, and $m_2$ are expressed as
\begin{equation}\label{eq2}
\left\{
\begin{array}{l}
m_0(x,y)=\frac{1}{2}g(x,y)[k_1(x,y)+k_2(x,y)] \\
m_1(x,y)=\frac{1}{2}g(x,y)[k_1(x,y)-k_2(x,y)]\cos2\theta(x,y) \\
m_2(x,y)=\frac{1}{2}g(x,y)[k_1(x,y)-k_2(x,y)]\sin2\theta(x,y)
\end{array}
\!.\right.
\end{equation}
Here, $g$ represents the gain of the photodetector, $k_1$ and $k_2$ represent the major and minor principal transmittances of the linear polarization filter, respectively, and $\theta$ represents the orientation of the linear polarization filter.

In the interpolation-based and filtering-based methods, the modulation parameters are regarded as periodically ideal values, expressed as~\cite{RN1047}
\begin{equation}\label{eq3}
\left\{
\begin{array}{l}
m_0(x,y)=1\\
m_1(x,y)=\frac{1}{2}[\cos(\pi x)+\cos(\pi y)]\\
m_2(x,y)=\frac{1}{2}[\cos(\pi x)-\cos(\pi y)]
\end{array}
\!,\right.
\end{equation}
corresponding to $g(x,y)=2$, $k_1(x,y)=1$, $k_2(x,y)=0$, and $\theta(x,y)$ takes $0$, $\pi/4$, $\pi/2$, or $3\pi/4$ according to the arrangement shown in Fig.~\ref{fig1}. In practice, due to the existence of the non-uniformity, the major and minor principal transmittances of the linear polarization filters and the gains and dark offsets of the photodetectors generally show individual differences, and the actual orientations of the linear polarization filters are also deviated from the designed orientations. To tackle the reconstruction errors and artifacts caused by the non-uniformity, these parameters need to be calibrated. The calibration of these parameters has been well discussed in Refs.~\cite{RN834,RN1720,RN1637}, therefore we consider $m_0$, $m_1$, $m_2$, and $d$ as known. Notice that here the term calibration refers to experimental processes for measuring the modulation and offset parameters, instead of numerical methods for correcting the non-uniform intensity responses in the DoFP image to ideal~\cite{RN834}. To normalize the modulation parameters and choose an orientation as the reference of 0$^{\circ}$, we apply the scaling and rotation transformation given in~\ref{app:A} to the modulation parameters. And since the influence caused by the non-uniformity of $d$ can be mitigated by simply subtracting $d$ from $i$, we will omit $d$ for brevity.

\begin{figure}[hbt!]
	\centering\includegraphics[width=\textwidth]{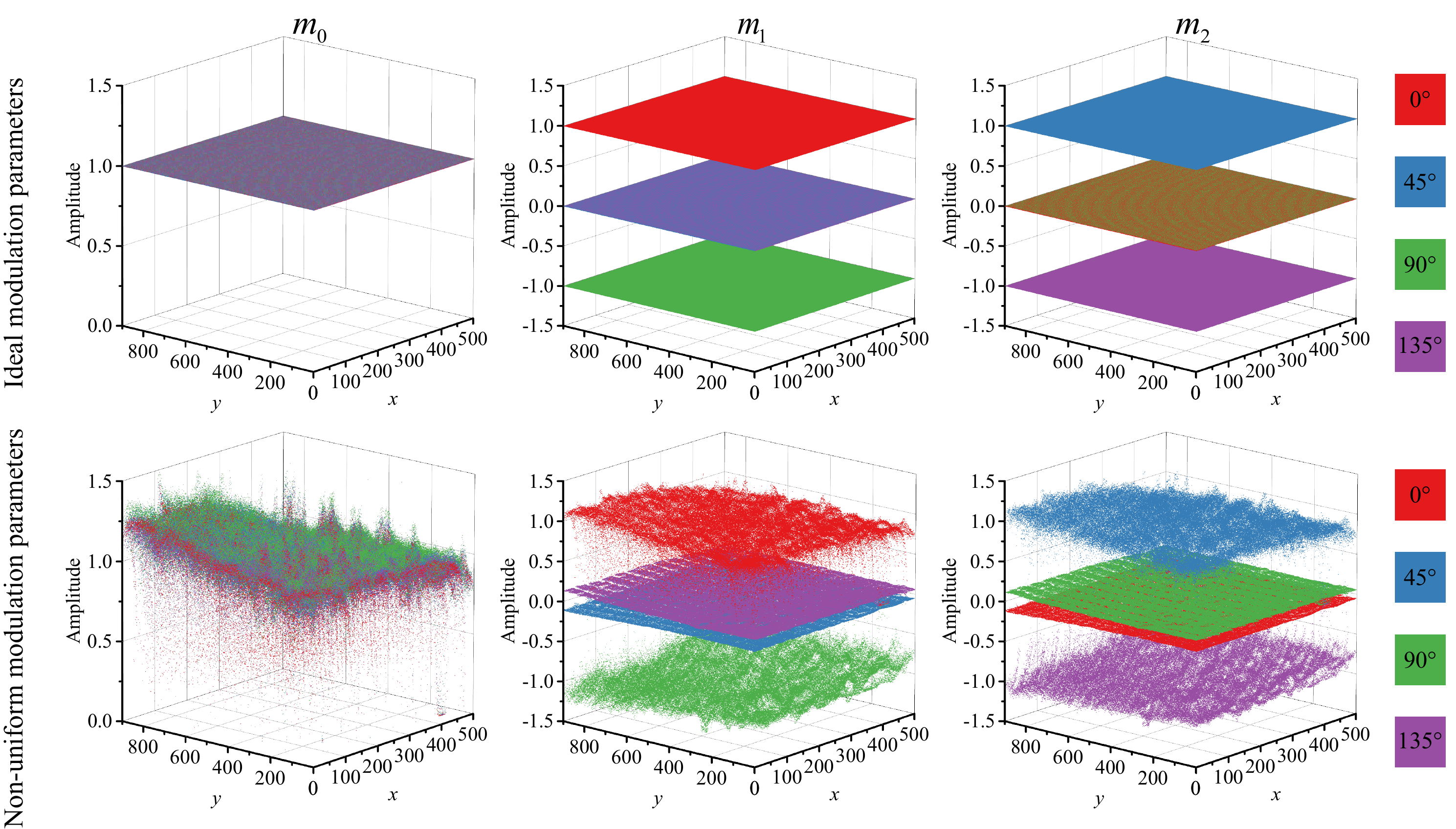}
	\caption{Scatterplots of the ideal and 
	non-uniform modulation parameters. The points are painted with four different colors according to the designed polarization orientations}
	\label{fig2}
\end{figure}

To generalize our discussions, we consider two sets of modulation parameters. One is the ideal modulation parameters given in Eq.~(\ref{eq3}). The scatterplots of the ideal modulation parameters are shown in the first row of Fig.~\ref{fig2}. We use the ideal modulation parameters to give the theoretical evaluations of the performance of the proposed methods, and explain the relationship between the proposed methods and the interpolation-based and filtering-based methods. The other is the non-uniform modulation parameters obtained from the calibration of our self-developed DoFP polarimeters. The second row of Fig.~\ref{fig2} shows the scatterplots of the non-uniform modulation parameters. And Table~\ref{tab1} gives the means and standard deviations of $g\cdot(k_1+k_2)$, $g\cdot(k_1-k_2)$, and $\theta$ in the non-uniform modulation parameters according to the designed polarization orientations. It can be seen that the non-uniform modulation parameters show strong non-uniformity deviations. We use the non-uniform modulation parameters to illustrate the abilities of the OLSM and SRM for mitigating the reconstruction errors and artifacts caused by the non-uniformity.

\begin{table}[]
	\centering
	\caption{Means and standard deviations of $g\cdot(k_1+k_2)$, $g\cdot(k_1-k_2)$, and $\theta$.}
	\resizebox{\textwidth}{!}{
	\begin{tabular}{lllllllll}
		\hline
		\multirow[b]{2}{*}{\begin{tabular}[c]{@{}l@{}}Polarization\\	orientation\end{tabular}} & \multicolumn{2}{l}{$g\cdot(k_1+k_2)$} &  & \multicolumn{2}{l}{$g\cdot(k_1-k_2)$} &  & \multicolumn{2}{l}{$\theta$} \\ \cline{2-3} \cline{5-6} \cline{8-9} 
		& Mean & \begin{tabular}[c]{@{}l@{}}Standard\\ deviation\end{tabular} &  & Mean & \begin{tabular}[c]{@{}l@{}}Standard\\ deviation\end{tabular} &  & Mean & \begin{tabular}[c]{@{}l@{}}Standard\\ deviation\end{tabular} \\ \hline
		0$^{\circ}$ & 0.9583 & 0.1818 &  & 0.8822 & 0.1766 &  & -2.4422$^{\circ}$ & 1.7530$^{\circ}$ \\
		45$^{\circ}$ & 1.0083 & 0.1288 &  & 0.9161 & 0.1252 &  & 46.9305$^{\circ}$ & 1.8916$^{\circ}$ \\
		90$^{\circ}$ & 1.0458 & 0.1319 &  & 0.9668 & 0.1286 &  & 87.8559$^{\circ}$ & 2.4745$^{\circ}$ \\
		135$^{\circ}$ & 0.9876 & 0.1314 &  & 0.9141 & 0.1277 &  & 137.5902$^{\circ}$ & 1.2284$^{\circ}$ \\ \hline
	\end{tabular}}
	\label{tab1}
\end{table}
\section{Stokes parameters reconstruction}\label{sec3}
Reconstructing the Stokes parameters requires dealing with the underdetermined system of linear equations composed by Eq.~(\ref{eq1}). This problem has infinite solutions. Generally, some prior constraints need to be introduced to find a desired solution. In this section, the OLSM and SRM are presented to solve the underdetermined system of linear equations by applying the local constant assumption and global smoothing assumption, respectively.
\subsection{Ordinary least-squares method}
For each $2\times2$ subset in the DoFP image, there are four equality constraints and twelve unknown Stokes parameters. By applying the local constant assumption that the Stokes parameters are constant functions in $2\times2$ subsets, the number of the unknowns in each $2\times2$ subset is reduced to three. Then the Stokes parameters can be determined according to the ordinary least-squares criterion, expressed as
\begin{equation}\label{eq4}
\begin{bmatrix}
	{{{\hat s}_0}(x + \tfrac{1}{2},y + \tfrac{1}{2})} \\ 
	{{{\hat s}_1}(x + \tfrac{1}{2},y + \tfrac{1}{2})} \\ 
	{{{\hat s}_2}(x + \tfrac{1}{2},y + \tfrac{1}{2})} 
\end{bmatrix} 
= 
\begin{bmatrix}
		{{m_0}(x,y)}&{{m_1}(x,y)}&{{m_2}(x,y)} \\ 
		{{m_0}(x,y + 1)}&{{m_1}(x,y + 1)}&{{m_2}(x,y + 1)} \\ 
		{{m_0}(x + 1,y)}&{{m_1}(x + 1,y)}&{{m_2}(x + 1,y)} \\ 
		{{m_0}(x + 1,y + 1)}&{{m_1}(x + 1,y + 1)}&{{m_2}(x + 1,y + 1)} 
\end{bmatrix}
^\dag
{\begin{bmatrix}
	{i(x,y)} \\ 
	{i(x,y + 1)} \\ 
	{i(x + 1,y)} \\ 
	{i(x + 1,y + 1)} 
\end{bmatrix}}.
\end{equation}
Here, ${\hat s}_0$, ${\hat s}_1$, and ${\hat s}_2$ represent the reconstructed Stokes parameters, and $[\bullet]^\dagger$ represents the pseudo-inverse of a matrix. The feasibility of using Eq.~(\ref{eq4}) to reconstruct the Stokes parameters has been demonstrated in Ref.~\cite{david2008unpolarized,RN1643,RN1637}. However, when substituting the ideal modulation parameters into Eq.~(\ref{eq4}), it can be found that the reconstruction results of Eq.~(\ref{eq4}) are equivalent to that of the nearest-neighbor interpolation-based method. This indicates that when facing the spatial variations of the Stokes parameters, Eq.~(\ref{eq4}) performs as poorly as the nearest-neighbor interpolation-based reconstruction.

Considering that each pixel is contained in four different 2$\times$2 subsets, we further apply a four-subset averaging strategy to obtain the final reconstruction results, expressed as

\begin{equation}\label{eq5}
{\hat s_k}(x,y)  = \tfrac{1}{4} [{{\hat s}_k}(x - \tfrac{1}{2},y - \tfrac{1}{2}) + {{\hat s}_k}(x - \tfrac{1}{2},y + \tfrac{1}{2}) + {{\hat s}_k}(x + \tfrac{1}{2},y - \tfrac{1}{2}) + {{\hat s}_k}(x + \tfrac{1}{2},y + \tfrac{1}{2})].
\end{equation}
Here, the subscript $k$ takes $0$, $1$, and $2$. The four-subset averaging strategy is also an optimization in the sense of the ordinary least-squares. With this strategy, when substituting the ideal modulation parameters, it can be found that the reconstruction results of the OLSM (Eqs.~(\ref{eq4}) and (\ref{eq5})) are  equivalent to that of the bilinear interpolation-based method.
\subsection{Smoothing regularization method}
Regularization is a commonly used optimization technique for solving underdetermined problems~\cite{RN1728,boyd2004convex}. We assume that the Stokes parameters are spatially smooth and apply the regularization technique to find the globally smoothing solution of the system of linear equations composed by Eq.~(\ref{eq1}). The Stokes parameters are determined by minimizing the objective function $L$ defined as
\begin{equation}\label{eq6}
\begin{split}
L({{\hat s}_0},{{\hat s}_1},{{\hat s}_2}) &= \sum\limits_{x,y} {{\left[ {{m_0}(x,y){{\hat s}_0}(x,y) + {m_1}(x,y){{\hat s}_1}(x,y) + {m_2}(x,y){{\hat s}_2}(x,y) - i(x,y)} \right]}^2}\\
&+ {\lambda _0}R({{\hat s}_0}) + {\lambda _1}R({\kappa _1}\tfrac{{{{\hat s}_1} + {{\hat s}_2}}}{2}) + {\lambda _2}R({\kappa _2}\tfrac{{{{\hat s}_1} - {{\hat s}_2}}}{2}).
\end{split}
\end{equation}
Here, the first term on the right side of Eq.~\ref{eq6} is the fidelity term used to penalize the deviation between the reconstructed Stokes parameters and the constraint of Eq.~(\ref{eq1}), $R(\hat s_0)$, $R(\kappa _1\tfrac{\hat s_1+\hat s_2}{2})$, and $R(\kappa _2\tfrac{\hat s_1-\hat s_2}{2})$ are the regularization terms, $R$ is the discrete thin-plate energy functional used to introduce the constraint of the spatial smoothness, defined as
\begin{equation}\label{eq7}
\begin{split}
R(f) &= \sum\limits_{x,y} {{{\left[ {f(x - 1,y) - 2f(x,y) + f(x + 1,y)} \right]}^2}} \\
&+ \sum\limits_{x,y} {{{\left[ {f(x,y - 1) - 2f(x,y) + f(x,y + 1)} \right]}^2}} \\
&+ 2\sum\limits_{x,y} {{{\left[ {f(x,y) - f(x + 1,y) - f(x,y + 1) + f(x + 1,y + 1)} \right]}^2}},
\end{split}
\end{equation}
$\lambda_0$, $\lambda_1$, and $\lambda_2$ are the regularization parameters used to control the weights of the regularization terms, and $\kappa_1$ and $\kappa_2$ are two parameters used to compensate the change of the relative weights between the fidelity term and regularization terms caused by the non-uniformity, defined as
\begin{equation}\label{eq8}
\left\{
\begin{array}{l}
{\kappa_1} = \frac{{\sum\limits_{x,y} {{m_1}(x,y)\cos (\pi x)}  + \sum\limits_{x,y} {{m_2}(x,y)\cos (\pi x)} }}{{\sum\limits_{x,y} {{m_0}(x,y)} }}\\
{\kappa_2} = \frac{{\sum\limits_{x,y} {{m_1}(x,y)\cos (\pi y)}  - \sum\limits_{x,y} {{m_2}(x,y)\cos (\pi y)} }}{{\sum\limits_{x,y} {{m_0}(x,y)} }}
\end{array}
\!.\right.
\end{equation}

The desired Stokes parameters should satisfy the Euler-Lagrange equation of the objective function, expressed as
\begin{equation}\label{eq9}
\left\{
\begin{array}{l}
\frac{{\partial L}}{{\partial {{\hat s}_0}}} = m_0^2{{\hat s}_0} + {m_0}{m_1}{{\hat s}_1} + {m_0}{m_2}{{\hat s}_2} - {m_0}i + {\lambda _0}{\nabla ^4}{{\hat s}_0} = 0 \\

\frac{{\partial L}}{{\partial {{\hat s}_1}}} = {m_0}{m_1}{{\hat s}_0} + m_1^2{{\hat s}_1} + {m_1}{m_2}{{\hat s}_2} - {m_1}i + \frac{1}{4}( {{\lambda _1}\kappa _1^2 + {\lambda _2}\kappa _2^2} ){\nabla ^4}{{\hat s}_1} + \frac{1}{4}( {{\lambda _1}\kappa _1^2 - {\lambda _2}\kappa _2^2} ){\nabla ^4}{{\hat s}_2} = 0 \\

\frac{{\partial L}}{{\partial {{\hat s}_2}}} = {m_0}{m_2}{{\hat s}_0} + {m_1}{m_2}{{\hat s}_1} + m_2^2{{\hat s}_2} - {m_2}i + \frac{1}{4}( {{\lambda _1}\kappa _1^2 - {\lambda _2}\kappa _2^2} ){\nabla ^4}{{\hat s}_1} + \frac{1}{4}( {{\lambda _1}\kappa _1^2 + {\lambda _2}\kappa _2^2} ){\nabla ^4}{{\hat s}_2} = 0 
\end{array}
\!.\right.
\end{equation}
Here, $\nabla ^4$ represents the discrete biharmonic operator, corresponding to the variation of the discrete thin-plate functional. The discrete biharmonic operator is implemented by the convolution operation
\begin{equation}\label{eq10}
{\nabla ^4}f = f * 
\begin{bmatrix}
	0&0&1&0&0 \\ 
	0&2&{ - 8}&2&0 \\ 
	1&{ - 8}&{20}&{ - 8}&1 \\ 
	0&2&{ - 8}&2&0 \\ 
	0&0&1&0&0 
\end{bmatrix}.
\end{equation}
Here, $*$ represents the convolution operation.

Equation~(\ref{eq9}) can be solved by gradient descent algorithms. We gave a MATLAB code implementation of the SRM in Ref.~\cite{srm}. Since the objective function is a convex function, the calculations can well converge to the global optimal solution that minimizes the objective function.

It is worth pointing out that the proposed methods and numerical calibration methods tackle the non-uniformity based on the same pixel model but have different purposes. The proposed methods aims to directly reconstruct the Stokes parameters, while the numerical calibration methods aims to correct the non-uniform intensity responses in the DoFP image. The relationship between them is further explained in~\ref{app:D}.
\section{Reconstruction error evaluations}\label{sec4}
In the DoFP modulation, the measurement information from different pixel coordinates is combined to reconstruct the Stoke parameters. Consequently, the spatial variations of the Stokes parameters will introduce reconstruction errors, which is significantly different from the other modulation techniques. 
In this section, the reconstruction errors of the OLSM and SRM for the different frequency components of $s_0$, $(s_1+s_2)/2$, and $(s_1-s_2)/2$ are evaluated. Firstly, Fourier analysis is applied to evaluate the reconstruction errors in the ideal case that the Stokes parameters are modulated by the ideal modulation parameters. Secondly, numerical simulations are applied to evaluate the reconstruction errors in the non-uniform case that the Stokes parameters are modulated by the non-uniform modulation parameters.
\subsection{Fourier analysis}
\begin{figure}[hbt!]
	\centering\includegraphics[width=8cm]{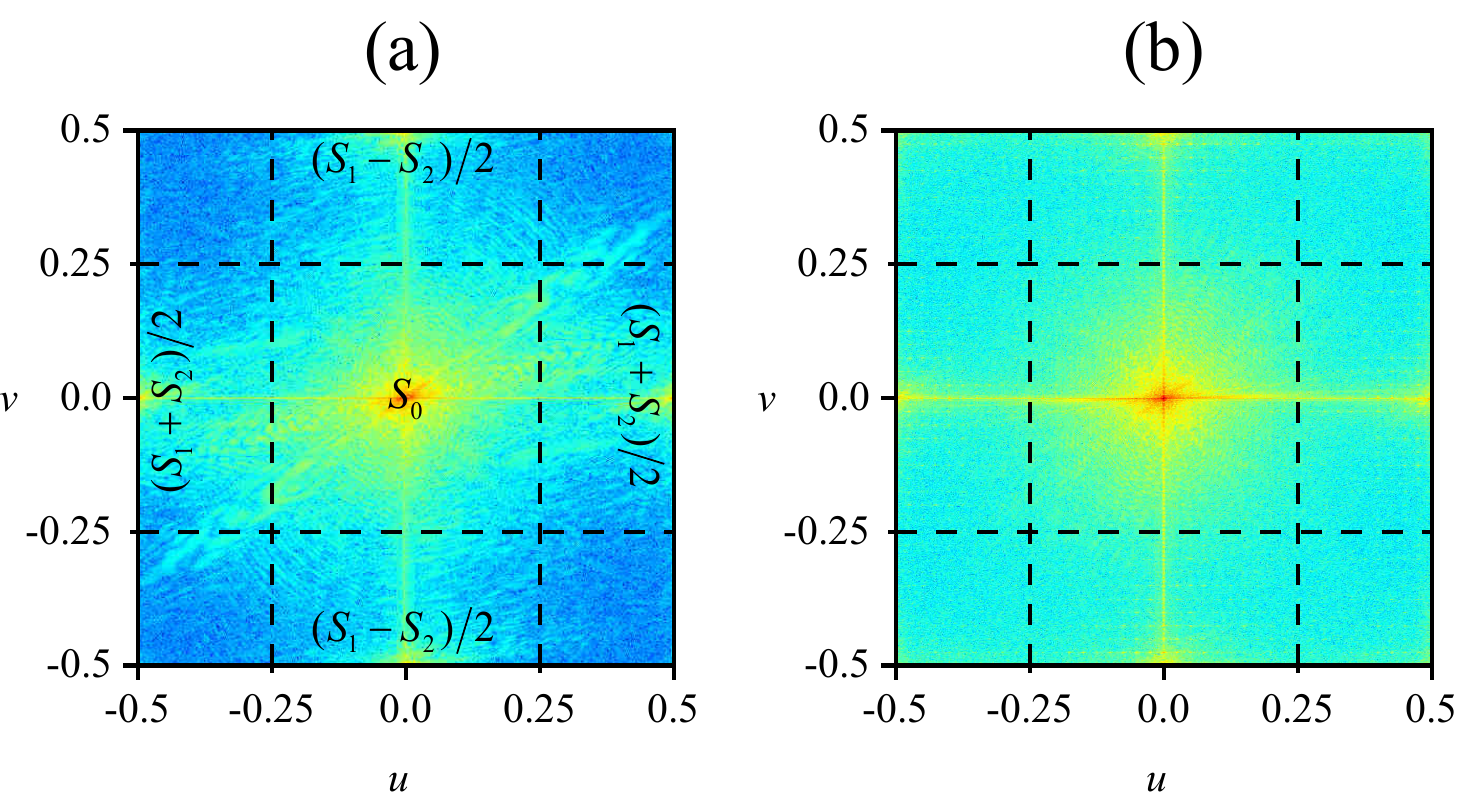}
	\caption{Examples of the log-scale spectrum of $i$. (a) The Stokes parameters are modulated by the ideal modulation parameters. (b) The Stokes parameters are modulated by the non-uniform modulation parameters.}
	\label{fig3}
\end{figure}

When the Stokes parameters are modulated by the ideal modulation parameters, substituting Eq.~(\ref{eq3}) into Eq.~(\ref{eq1}), the discrete Fourier transform (DFT) of $i$ is expressed as
\begin{equation}\label{eq11}
I(u,v) = {S_0}(u,v) + \tfrac{1}{2}[{S_1}(u + \tfrac{1}{2},v) + {S_2}(u + \tfrac{1}{2},v)] + \tfrac{1}{2}[{S_1}(u,v + \tfrac{1}{2}) - {S_2}(u,v + \tfrac{1}{2})].
\end{equation}
Here, $u$ and $v$ represent the horizontal and vertical frequency coordinates, respectively, and the uppercase symbols represent the DFT of the corresponding lowercase symbols. Equation.~(\ref{eq11}) indicates that the frequency components of $s_0$ are located in the center region of the spectrum of $i$, while the frequency components of $(s_1+s_2)/2$ and $(s_1-s_2)/2$ are shifted into the horizontal and vertical border regions, respectively. Figure.~\ref{fig3}(a) gives an example of the spectrum of $i$ where the Stokes parameters are modulated by the ideal modulation parameters (see Section~\ref{sec6}). It can be seen that there are noticeable carrier peaks at the frequency coordinates $(0,0)$, $(\pm\frac{1}{2},0)$, and $(0,\pm\frac{1}{2})$.

\begin{figure}[hbt!]
	\centering\includegraphics[width=\textwidth]{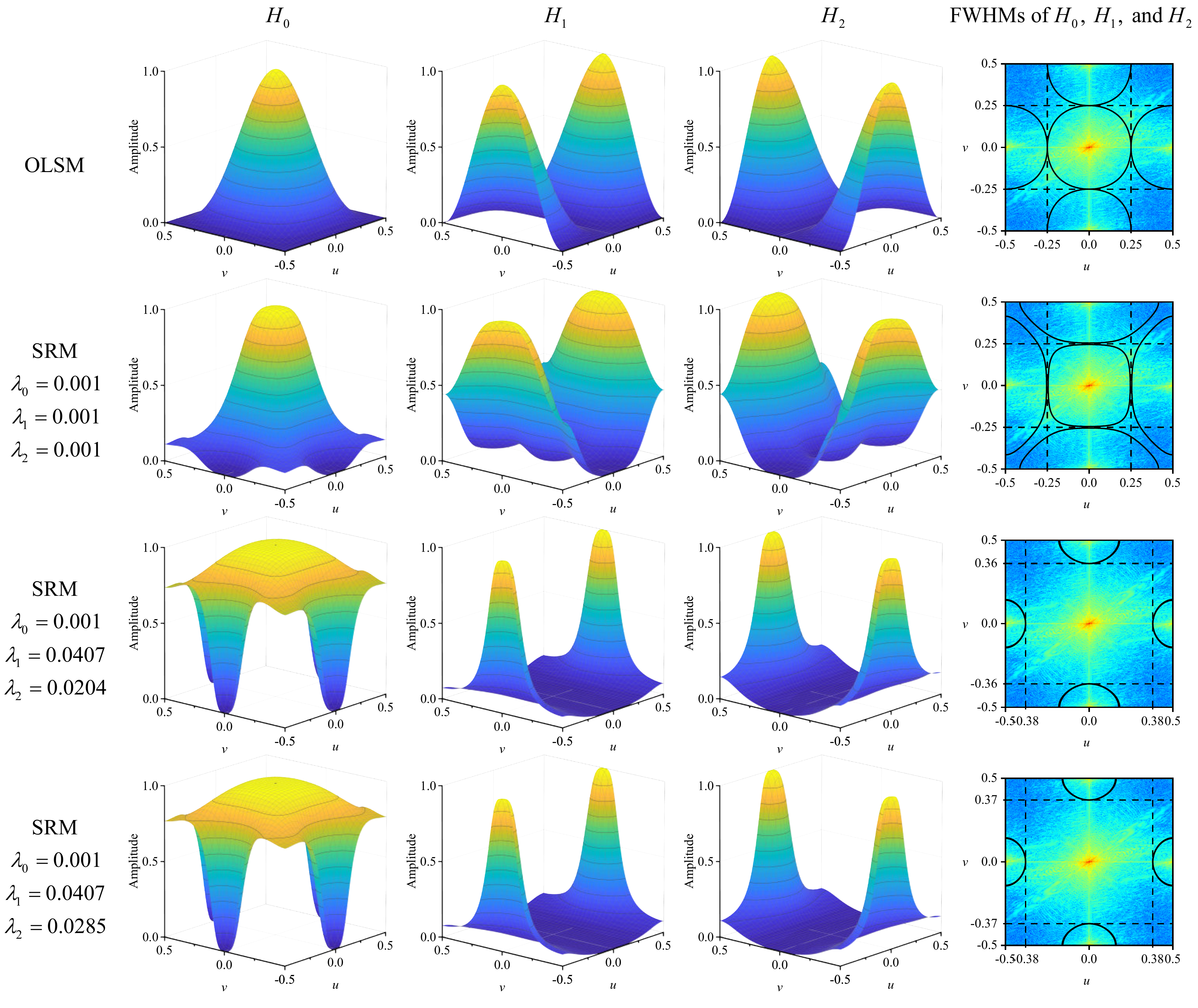}
	\caption{Filter transfer functions and their FWHMs.}
	\label{fig4}
\end{figure}

The distribution of the frequency components of $s_0$, $(s_1+s_2)/2$, and $(s_1-s_2)/2$ allows us to reconstruct the Stokes parameters through frequency domain filtering, expressed as
\begin{equation}\label{eq12}
\left\{
\begin{array}{l}
{{\hat S}_0}(u,v) = I(u,v) \cdot {H_0}(u,v)\\
\frac{1}{2}[{{\hat S}_1}(u + \tfrac{1}{2},v) + {{\hat S}_2}(u + \tfrac{1}{2},v)] = I(u,v) \cdot {H_1}(u,v)\\
\frac{1}{2}[{{\hat S}_1}(u,v + \tfrac{1}{2}) - {{\hat S}_2}(u,v + \tfrac{1}{2})] = I(u,v) \cdot {H_2}(u,v) 
\end{array}
\!.\right.
\end{equation}
Here, $H_0$, $H_1$, and $H_2$ represent the filter transfer functions which aim to retrieve the frequency components of $s_0$, $(s_1+s_2)/2$, and $(s_1-s_2)/2$, respectively. In the filtering-based methods, the filter transfer functions are constructed by window functions such as Hamming, Gaussian, and Planck-taper window functions~\cite{RN1047,RN1665,RN1051}. In fact, ignoring the differences around the image boundaries of the reconstruction results, we find that the convolutional interpolation-based methods, OLSM, and SRM have equivalent implementations in the frequency domain. The filter transfer functions of the mainstream convolutional interpolation-based methods are given in~\ref{app:B}, including the nearest-neighbor, bilinear, bicubic, and natural bicubic spline interpolation-based methods. For the OLSM, substituting Eq.~(\ref{eq3}) into Eqs.~(\ref{eq4}) and (\ref{eq5}), and performing the DFT, after some simplifications, we have
\begin{equation}\label{eq13}
\left\{
\begin{array}{l}
{H_0}(u,v) = \frac{1}{4}[ {1 + \cos (2\pi u)} ][ {1 + \cos (2\pi v)} ] \\
{H_1}(u,v) = \frac{1}{4}[ {1 + \cos (2\pi u + \pi )} ][ {1 + \cos (2\pi v)} ] \\
{H_2}(u,v) = \frac{1}{4}[ {1 + \cos (2\pi u)} ][ {1 + \cos (2\pi v + \pi )} ]
\end{array} 
\!.\right.
\end{equation}
The first row of Fig.~\ref{fig4} shows the filter transfer functions of the OLSM and their full-widths-at-half-maximum (FWHMs). Notice that the OLSM is equivalent to the bilinear interpolation-based method in this case. For the SRM, substituting Eq.~(\ref{eq3}) into Eq.~(\ref{eq9}), and performing the DFT, after the same simplifications, we have
\begin{equation}\label{eq14}
\left\{
\begin{array}{l}
{H_0}(u,v) = {\left[ {1 + {\lambda _0}G(u,v) + \frac{{{\lambda _0}G(u,v)}}{{{\lambda _1}G(u + \tfrac{1}{2},v)}} + \frac{{{\lambda _0}G(u,v)}}{{{\lambda _2}G(u,v + \tfrac{1}{2})}}} \right]^{ - 1}} \\
{H_1}(u,v) = {\left[ {1 + {\lambda _1}G(u + \tfrac{1}{2},v) + \frac{{{\lambda _1}G(u + \tfrac{1}{2},v)}}{{{\lambda _0}G(u,v)}} + \frac{{{\lambda _1}G(u + \tfrac{1}{2},v)}}{{{\lambda _2}G(u,v + \tfrac{1}{2})}}} \right]^{ - 1}} \\
{H_2}(u,v) = {\left[ {1 + {\lambda _2}G(u,v + \tfrac{1}{2}) + \frac{{{\lambda _2}G(u,v + \tfrac{1}{2})}}{{{\lambda _0}G(u,v)}} + \frac{{{\lambda _2}G(u,v + \tfrac{1}{2})}}{{{\lambda _1}G(u + \tfrac{1}{2},v)}}} \right]^{ - 1}} 
\end{array}
\!.\right.
\end{equation}
Here, $G$ is the DFT of the discrete biharmonic operator, expressed as
\begin{equation}\label{eq15}
\begin{split}
G(u,v) &= 20 - 16\cos (2\pi u) - 16\cos (2\pi v) + 4\cos (2\pi u + 2\pi v) \\
& + 4\cos (2\pi u - 2\pi v) + 2\cos (4\pi u) + 2\cos (4\pi v).
\end{split}
\end{equation}
The filter transfer functions of the SRM are adjustable by changing the values of the regularization parameters. For arbitrary $\lambda_1/\lambda_0$ and $\lambda_2/\lambda_0$, when $\lambda_0$ tends to 0, these filter transfer functions satisfy
\begin{equation}\label{eq16}
\left\{
\begin{array}{l}
{H_0}( \pm \frac{1}{\pi }\arctan \left( {{{(\frac{{{\lambda _1}}}{{{\lambda _0}}})}^{\frac{1}{4}}}} \right),0) = {H_1}( \pm \frac{1}{\pi }\arctan \left( {{{(\frac{{{\lambda _1}}}{{{\lambda _0}}})}^{\frac{1}{4}}}} \right),0) = 0.5 \\
{H_0}(0, \pm \frac{1}{\pi }\arctan \left( {{{(\frac{{{\lambda _2}}}{{{\lambda _0}}})}^{\frac{1}{4}}}} \right)) = {H_2}(0, \pm \frac{1}{\pi }\arctan \left( {{{(\frac{{{\lambda _2}}}{{{\lambda _0}}})}^{\frac{1}{4}}}} \right)) = 0.5 
\end{array}
\!.\right.
\end{equation}
The second, third, and fourth rows of Fig.~\ref{fig4} show the filter transfer functions of the SRM and their FWHMs with the regularization parameters chosen as "$\lambda_0=0.001$, $\lambda_1=0.001$, $\lambda_2=0.001$",  "$\lambda_0=0.001$, $\lambda_1=0.0407$, $\lambda_2=0.0204$", and "$\lambda_0=0.001$, $\lambda_1=0.0407$, $\lambda_2=0.0285$", respectively. It can be seen that the FWHMs along $u=0$ and $v=0$ in these examples are well consistent with Eq.~(\ref{eq16}).

In the perspective of sampling and reconstruction, the reconstruction errors come from the pre-aliasing occurring in the sampling process and the post-aliasing occurring in the reconstruction process. The pre-aliasing refers to the overlap of the frequency components of $s_0$, $(s_1+s_2)/2$, and $(s_1-s_2)/2$. Sufficient sampling is a necessary prerequisite to apply the DoFP technique, otherwise the strong pre-aliasing will always cause large reconstruction errors, which means the failure of the measurement. It has been show in Ref.~\cite{RN1047} that if the sampling can make the frequency components of $s_0$, $(s_1+s_2)/2$, and $(s_1-s_2)/2$ satisfy a band-limit condition, theoretically the Stokes parameters can be perfectly reconstructed without errors. Practically, the pre-aliasing is hard to avoid, while for most polarization imaging targets, the main energy of $s_0$, $(s_1+s_2)/2$, and $(s_1-s_2)/2$ to be measured are concentrated in their low-frequency components, and thus the pre-aliasing reconstruction errors are usually small.

The post-aliasing refers to the misscategorization of the frequency components of $s_0$, $(s_1+s_2)/2$, and $(s_1-s_2)/2$. For the nearest-neighbor interpolation-based method, its filter transfer functions cannot well attenuate the undesired frequency components, and therefore usually result in large post-aliasing reconstruction errors. For the OLSM and the bilinear, bicubic, and natural bicubic spline interpolation-based methods, their filter transfer functions indicate that only when the bandwidths occupied by the frequency components of $s_0$, $(s_1+s_2)/2$, and $(s_1-s_2)/2$ do not exceed 0.5 cycles per pixel, can the Stokes parameters be reconstructed with less post-aliasing reconstruction errors. For the SRM and filtering-based methods, their filter transfer functions are adjustable by changing regularization parameters and window function parameters, respectively. Considering that the practical bandwidths occupied by the frequency components of $s_0$, $(s_1+s_2)/2$, and $(s_1-s_2)/2$ are different for different polarization imaging targets, the adjustable filter transfer functions make these methods have more powerful abilities to reduce the post-aliasing reconstruction errors. Comparing the filter transfer functions of the SLM and the filter transfer functions constructed by the window functions used in previous studies, a noticeable difference is that the frequency responses of the filter transfer functions of the SRM for the frequency components of $s_0$, $(s_1+s_2)/2$, and $(s_1-s_2)/2$ are anisotropic. It can be seen from Fig.~\ref{fig3}(a) that the probabilities that the pre-aliasing occurs along different directions are different. This indicates that the anisotropic frequency responses can better reduce the reconstruction errors.
\subsection{Numerical simulations}
Figure.~\ref{fig3}(b) gives an example of the spectrum of $i$ where the Stokes parameters are modulated by the non-uniform modulation parameters (see Section~\ref{sec6}). There are still noticeable carrier peaks at the frequency coordinates $(0,0)$, $(\pm\frac{1}{2},0)$, and $(0,\pm\frac{1}{2})$. However, due to the existence of the non-uniformity, the spectrum is disturbed. In this case, if the ideal modulation parameters are still used in the reconstruction, according to the Fourier analysis, the reconstruction results are inevitably influenced by the non-uniformity. However, combining with the non-uniform modulation parameters, the OLSM and SRM are able to construct the filters with space-varying filter kernels to deal with the non-uniformity.

\begin{figure}[hbt!]
	\centering\includegraphics[width=\textwidth]{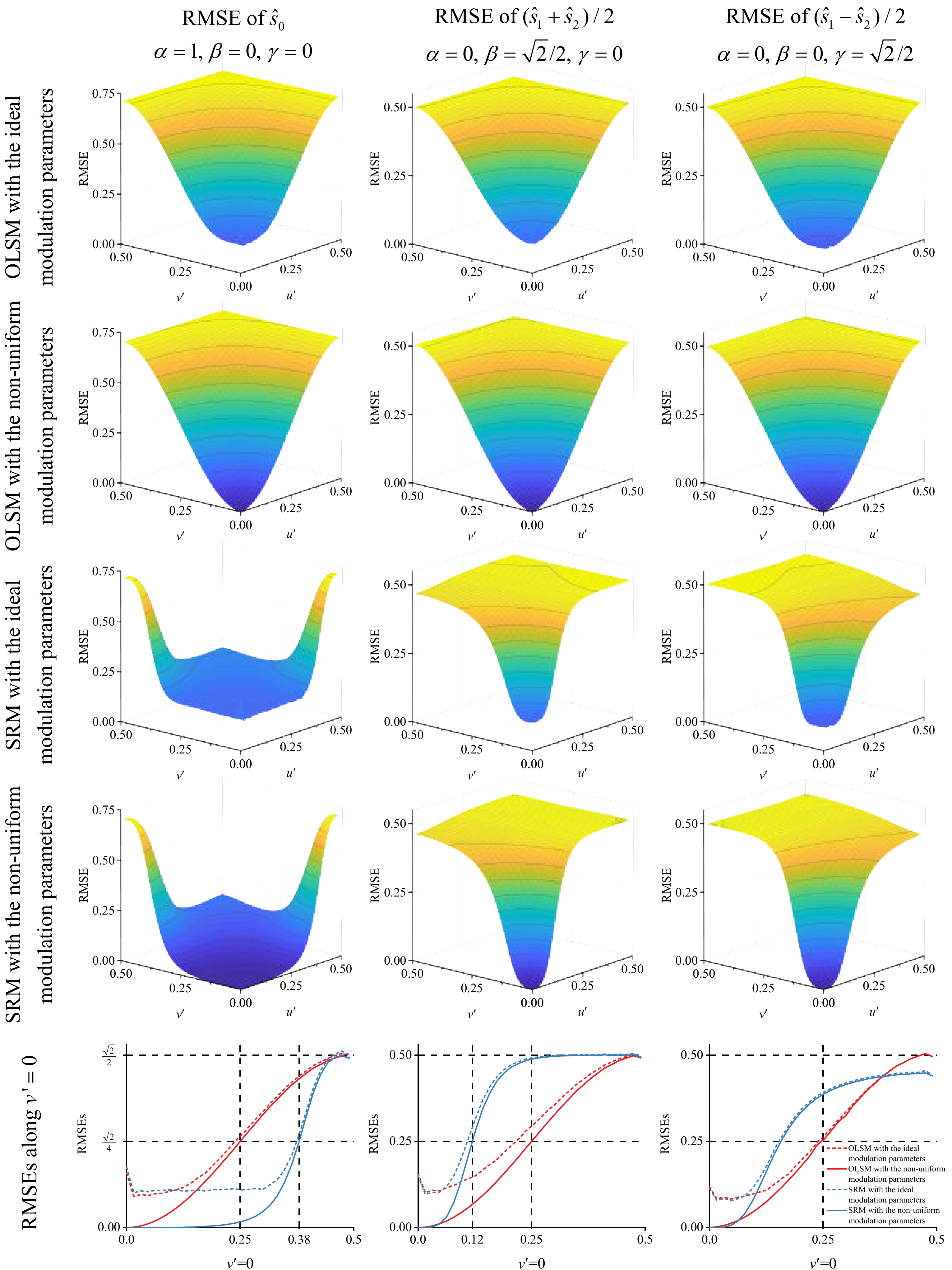}
	\caption{RMSEs of the reconstructed cosine patterns.}
	\label{fig5}
\end{figure}
Since the non-uniform modulation parameters are no longer periodic, we used numerical simulations to evaluate the reconstruction errors of the OLSM and SRM for the different frequency components of $s_0$, $(s_1+s_2)/2$, and $(s_1-s_2)/2$ in the non-uniform case. Firstly, we generated the cosine patterns of $s_0$, $(s_1+s_2)/2$, and $(s_1-s_2)/2$ by setting "$\alpha=1$, $\beta=0$, $\gamma=0$", "$\alpha=0$, $\beta=\sqrt{2}/2$, $\gamma=0$", and "$\alpha=0$, $\beta=0$, $\gamma=\sqrt{2}/2$", respectively, in the following equation
\begin{equation}\label{eq17}
\left\{
\begin{array}{l}
s_0(x,y) = 1 + \alpha \cos (2\pi u'x + 2\pi v'y) \\
\frac{1}{2}[{s_1(x,y) + s_2(x,y)}] = \beta \cos (2\pi u'x + 2\pi v'y) \\
\frac{1}{2}[{s_1(x,y) - s_2(x,y)}] = \gamma \cos (2\pi u'x + 2\pi v'y)
\end{array}
\!.\right.
\end{equation}
Here, $u'$ and $v'$ are the horizontal and vertical frequencies varying from 0 to 0.5 cycles per pixel. The choices of $\alpha$, $\beta$, and $\gamma$ make the generated cosine patterns satisfy the physical constraint $0 \leq \sqrt {{s_1}^2 + {s_2}^2}/{s_0} \leq 1$. Secondly, we substituted the generated cosine patterns and the non-uniform modulation parameters into Eq.~(\ref{eq1}) to generate the DoFP images. Lastly, we used the OLSM and SRM combined with the ideal and non-uniform modulation parameters to reconstruct the Stokes parameters from the generated DoFP images. The regularization parameters used in the numerical simulations were chosen as "$\lambda_0=0.001$, $\lambda_1=0.0407$, $\lambda_2=0.0285$" as example. The first and second rows of Fig.~\ref{fig5} show the root mean square errors (RMSEs) of the cosine patterns reconstructed by the OLSM combined with the ideal and non-uniform modulation parameters, respectively, the third and fourth rows of Fig.~\ref{fig5} show the RMSEs of the cosine patterns reconstructed by the SRM combined with the ideal and non-uniform modulation parameters, respectively, and the last row of Fig.~\ref{fig5} shows the comparisons of the above RMSEs along $v'=0$.

For the both methods, it can be seen that the high-frequency components of $s_0$, $(s_1+s_2)/2$, and $(s_1-s_2)/2$ tend to have large reconstruction errors. Therefore, it is still necessary to ensure sufficient sampling to reduce the energy in the high-frequency components. When the ideal modulation parameters are used, the low-frequency components of $s_0$, $(s_1+s_2)/2$, and $(s_1-s_2)/2$ have noticeable reconstruction errors, but when the non-uniform modulation parameters are used, the reconstruction errors for the low-frequency components of $s_0$, $(s_1+s_2)/2$, and $(s_1-s_2)/2$ are reduced, and the RMSEs reach zero for the zero-frequency components of $s_0$, $(s_1+s_2)/2$, and $(s_1-s_2)/2$. This indicates that when the non-uniformity is correctly characterized, the space-varying filter kernels constructed by the OLSM and SRM can reduce the reconstruction errors caused by the non-uniformity. It is worth pointing out that when the non-uniform modulation parameters are used, the RMSEs in Fig.~\ref{fig5} show exactly opposite trends compared with the frequency responses of the filter transfer functions in Fig.~\ref{fig4}. This phenomenon indicates that the overall frequency responses of the space-varying filter kernels for the frequency components of $s_0$, $(s_1+s_2)/2$, and $(s_1-s_2)/2$ are consistent with that of the filter transfer functions. The reason for this phenomenon is that the non-uniform modulation parameters are still close to the ideal modulation parameters in average sense.

\section{Choice of the regularization parameters}\label{sec5}
In this section, we give some rules of thumb for choosing suitable regularization parameters in the SRM. We represente the regularization parameters as $\lambda_0$, $\lambda_1/\lambda_0$, and $\lambda_2/\lambda_0$ to discuss their choices.

Keeping $\lambda_1/\lambda_0$ and $\lambda_2/\lambda_0$ invariant, $\lambda_0$ determines the overall weight of the regularization terms. The major factor influencing the choice of $\lambda_0$ is the level of the additional Gaussian noise. However, in this paper, the DoFP images are considered to be noiseless, so $\lambda_0$ needs to be set to a small value so that the reconstruction results can meet the constraint of Eq.~(\ref{eq1}). Empirically, $\lambda_0$ is chosen as 0.001, smaller values will not cause significant changes of the reconstruction results.

\begin{figure}[hbt!]
	\centering\includegraphics[width=8cm]{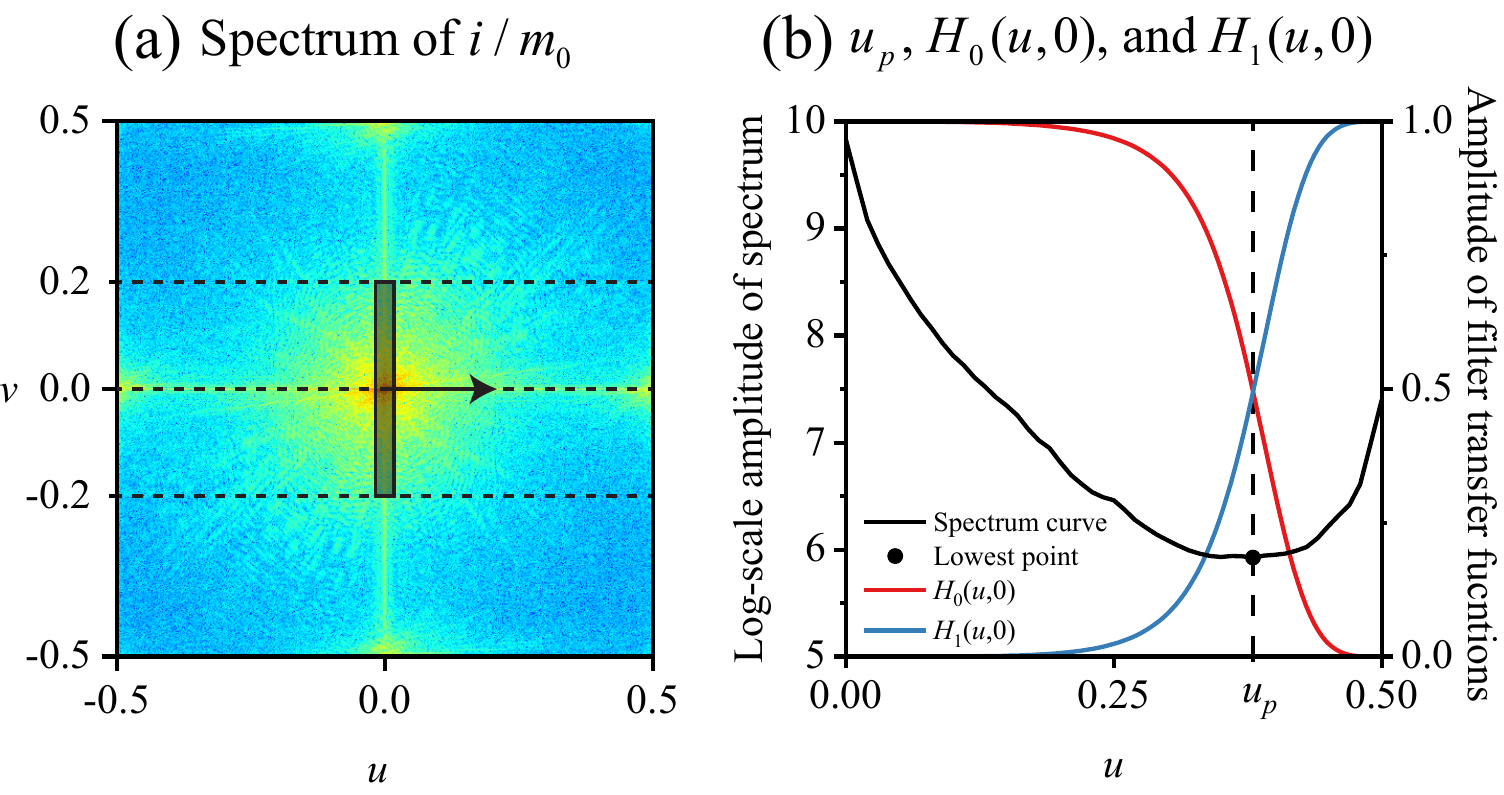}
	\caption{Strategy for choosing $\lambda_1/\lambda_0$. (a) Spectrum of $i/m_0$. A spectrum curve can be obtained by sliding a rectangular window horizontally and averaging the spectrum in the rectangular window. (b) $\lambda_1/\lambda_0$ is chosen to make the overall frequency responses of the filters for reconstructing $s_0$ and $(s_1+s_2)/2$ can be approximately equal at the frequency coordinate $(\pm u_p,0)$. $H_0$ and $H_1$ are used to represent the approximations of the overall frequency responses of the filters for reconstructing $s_0$ and $(s_1+s_2)/2$, respectively.}
	\label{fig6}
\end{figure}

$\lambda_1/\lambda_0$ determines the relative smoothness between the reconstructed $s_0$ and $(s_1+s_2)/2$. According to the evaluations in Section~\ref{sec4}, we need to choose a suitable value of $\lambda_1/\lambda_0$ so that the filters constructed by the SRM can well categorize the frequency components of $s_0$ and $(s_1+s_2)/2$. Our strategy for choosing $\lambda_1/\lambda_0$ is illustrated in Fig.~\ref{fig6}. As shown in Fig.~\ref{fig6}(a), to reduce the non-uniformity disturbances, the spectrum of $i/m_0$ is used for choosing $\lambda_1/\lambda_0$. By sliding the rectangular window whose center is located at the horizontal line $v=0$ and averaging the spectrum in the rectangular window, a spectrum curve can be obtained. Empirically, we choose the size of the rectangular window as $0.02\times0.4$. The frequency coordinate of the lowest point $u_p$ in the spectrum curve represents our estimate of the relative bandwidth occupied by the frequency components of $s_0$ and $(s_1+s_2)/2$ in the horizontal direction. As illustrated in Fig.~\ref{fig6}(b), we determine $\lambda_1/\lambda_0$ through the formula
\begin{equation}\label{eq18}
\frac{{{\lambda _1}}}{{{\lambda _0}}} = {\left[ {\tan (\pi {u_p})} \right]^4}.
\end{equation}
so that the overall frequency responses of the filters for reconstructing $s_0$ and $(s_1+s_2)/2$ can be approximately equal at the frequency coordinate $(\pm u_p,0)$.
Likewise, $\lambda_2/\lambda_0$ can be chosen by a similar strategy. In practice, inappropriate choices of $\lambda_1/\lambda_0$ and $\lambda_2/\lambda_0$ usually result in significantly serrated artifacts in the reconstructed results. If the above strategy fails, the regularization parameters can still be suitably chosen through visual inspection.

\section{Experiments}\label{sec6}
Two experiments were performed to evaluate and compare the performance of the OLSM, SRM, bicubic interpolation-based, Newton's polynomial interpolation-based method~\cite{RN1652}, and Planck-Taper window filtering-based methods. The polarization imaging target in the experiments is a model car. A DoT polarimeter and the self-developed DoFP polarimeter were used to capture the polarization information. Three measures were implemented to ensure sufficient sampling, including making the target occupy a large field of view, defocusing the target slightly, and reducing the lens aperture. 
The images recorded by the polarimeters were captured 100 times and averaged to decouple the effects of noise.

\begin{figure}[hbt!]
	\centering\includegraphics[width=\textwidth]{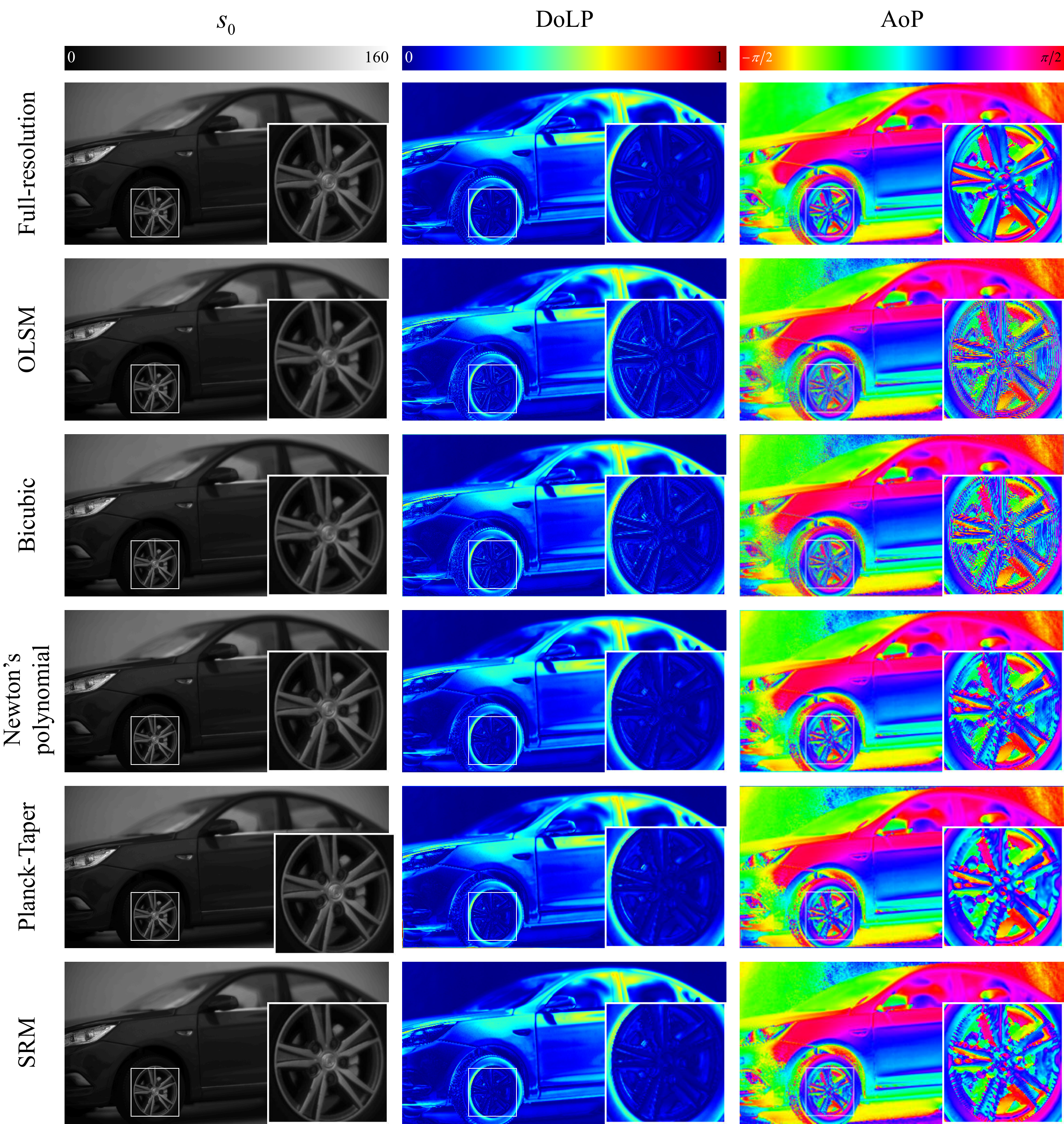}
	\caption{True and reconstructed $s_0$, DoLP, and AoP. The true Images in the white boxes show the magnification of local regions}
	\label{fig7}
\end{figure}

The DoT polarimeter is composed by a monochrome CCD camera (FLIR GS3-U3-15S5M-C) and a linear polarization filter (THORLABS WP25M-VIS) placed on a motorized rotating platform (THORLABS PRM1/MZ8). The DoT polarimeter captured 0$^{\circ}$, 45$^{\circ}$, 90$^{\circ}$, and 135$^{\circ}$ polarization images. These polarization images were used to calculate the full-resolution Stokes parameters, DoLP, and AoP. The results are regarded as the true values. By re-sampling the four polarization images, a DoFP image was synthesized. Benefiting from the high positioning accuracy of the rotating platform and the uniform performance of the camera and linear polarization filter, the Stokes parameters in the synthesized DoFP image can be regarded as being modulated by the ideal modulation parameters. The spectrum of the synthesized DoFP image is given in Fig.~\ref{fig3}(a). It can be seen that the frequency components of $s_0$ occupy more sized bandwidth than these of $(s_1+s_2)/2$ and $(s_1-s_2)/2$. As a matter of fact, this is a common phenomenon which has been shown in Ref.~\cite{RN1652,RN1051}. The OLSM, bicubic interpolation-based method, Newton's polynomial interpolation-based method, Planck-Taper window filtering-based method, and SRM were used to reconstruct the Stokes parameters, DoLP, and AoP from the synthesized DoFP image. The window function parameters of the Planck-Taper window functions are given in~\ref{app:C}. Since there is no good method for choosing suitable window function parameters, the best choice minimizing the RMSEs of $\hat s_0$, $(\hat s_1+ \hat s_2)/2$, and $(\hat s_1-\hat s_2)/2$ was used. The regularization parameters used in the SRM were chosen as "$\lambda_0=0.001$, $\lambda_1=0.0407$, $\lambda_2=0.0204$" according to the discussions given in Section~\ref{sec5}. Figure.~\ref{fig7} shows the true and reconstructed $s_0$, DoLP, and AoP. $s_0$ ranges from 0 to 255, while the range of the grayscale bar in Fig.~\ref{fig7} is set to $[0,160]$ to increase the image contrast. The most noticeable differences are marked in the white boxes. $s_0$ reconstructed by the OLSM and bicubic interpolation-based method are blurred, and the reconstructed DoLP and AoP show obviously serrated artifacts. The main reason for these phenomena is that the frequency components of $s_0$ that more than $\pm$ 0.25 cycles per pixel are incorrectly categorized. The Newton's polynomial interpolation-based method is an edge-preserved reconstruction method. It can be seen that the reconstructed $s_0$ shows more details, and the reconstructed DoLP image avoids the serrated artifacts. However, due to the wrong edge discriminations, the reconstructed AoP image has "$\times$" artifacts. The reconstructed results of the Planck-Taper window filtering-based method and SRM all show good visual effects. This benefits from the suitably choices of the window function parameters and regularization parameters.
Table~\ref{tab2} gives the RMSEs and normalized RMSEs of the reconstructed results. The normalized RMSEs are defined as the ratio of the RMSE to the difference between the maximum and minimum value of the true value. All the reconstruction methods can achieve relatively small reconstruction errors, which illustrates the effectiveness of the DoFP technique. Comparing the relative reconstruction errors of the different reconstruction items, it can be seen that the relative reconstruction errors of the AoP images are larger than that of the others. This is consistent with the results given in Ref.~\cite{RN1593}. The SRM shows minimal reconstruction errors for most reconstruction results. Meanwhile, due to the anisotropic frequency responses, the SRM shows smaller reconstruction errors than the Planck-Taper window filtering-based method for all the reconstruction results.

\begin{table}[]
	\centering
	\caption{RMSEs (\textit{normalized RMSEs}) of reconstructed $s_0$, $s_1$, $s_2$, DoLP, and AoP.}
	\resizebox{\textwidth}{!}{
	\begin{tabular}{llllll}
		\hline
		Reconstruction method & $s_0$ & $s_1$ & $s_2$ & DoLP & AoP \\ \hline
		\multirow{2}{*}{OLSM} & 0.67075 & 0.57365 & 0.56399 & 0.022871 & 0.20427 \\
		& (\textit{0.425\%}) & (\textit{1.188\%}) & (\textit{1.358\%}) & (\textit{3.210\%}) & (\textit{6.504\%}) \\
		\multirow{2}{*}{Bicubic} & 0.58439 & 0.53396 & 0.52278 & 0.020552 & 0.19243 \\
		& (\textit{0.371\%}) & (\textit{1.106\%}) & (\textit{1.259\%}) & (\textit{2.885\%}) & (\textit{6.125\%}) \\
		\multirow{2}{*}{Newton's polynomial} & 0.26618 & 0.27781 & 0.27062 & \textbf{0.015017} & 0.13984 \\
		& (\textit{0.169\%}) & (\textit{0.575\%}) & (\textit{0.652\%}) & (\textit{\textbf{2.108\%}}) & (\textit{4.451\%}) \\
		\multirow{2}{*}{Planck-Taper} & 0.31674 & 0.26161 & 0.25777 & 0.016486 & 0.13862 \\
		& (\textit{0.201\%}) & (\textit{0.542\%}) & (\textit{0.621\%}) & (\textit{2.314\%}) & (\textit{4.413\%}) \\
		\multirow{2}{*}{SRM} & \textbf{0.25188} & \textbf{0.25569} & \textbf{0.24729} & 0.015096 & \textbf{0.13747} \\
		&(\textit{\textbf{0.160\%}}) & (\textit{\textbf{0.530\%}}) & (\textit{\textbf{0.595\%}}) & (\textit{2.119\%}) & (\textit{\textbf{4.375\%}})\\
		\hline
	\end{tabular}}
	\label{tab2}
\end{table}     

\begin{figure}[hbt!]
	\centering\includegraphics[width=\textwidth]{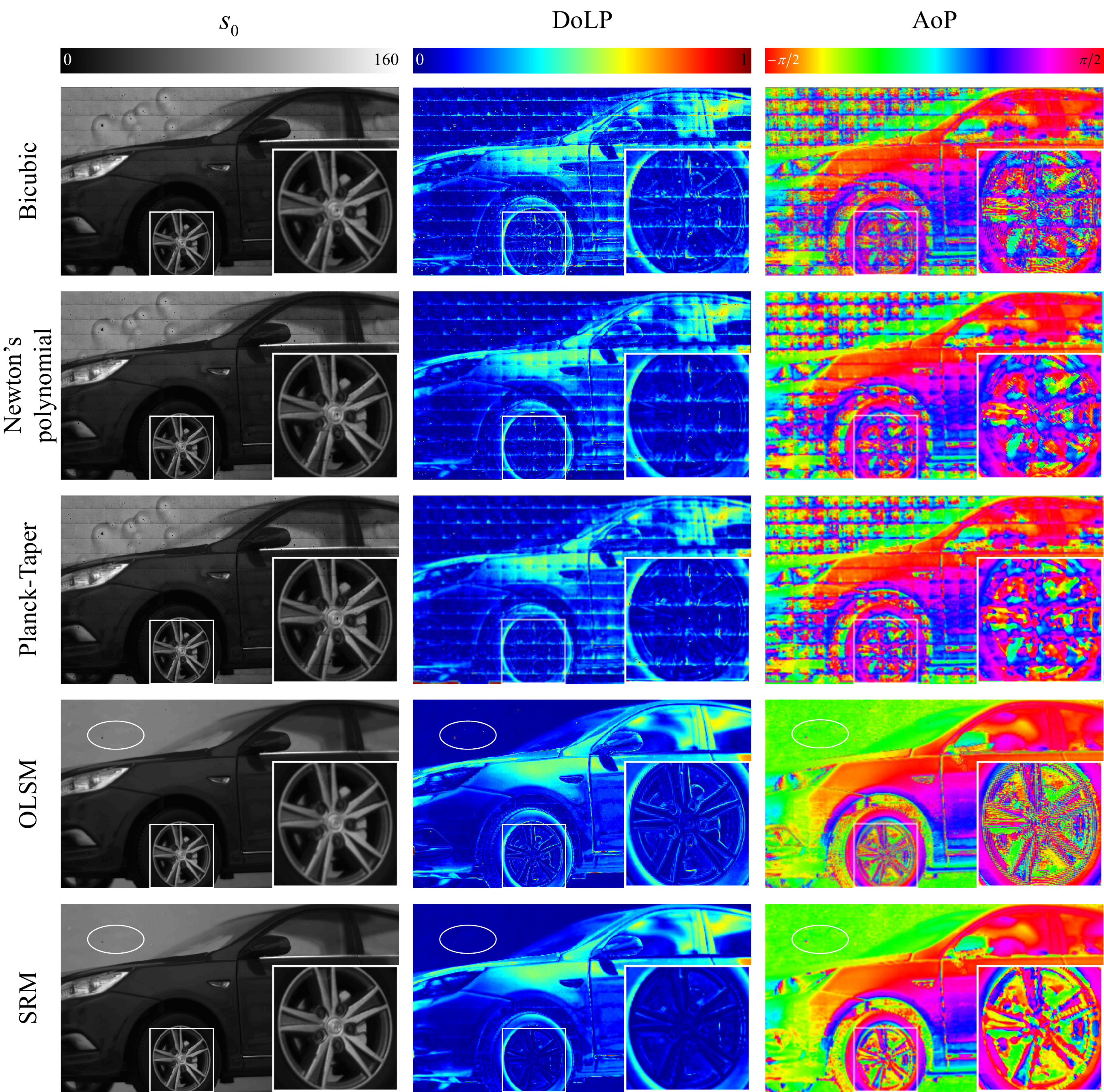}
	\caption{Reconstructed $s_0$, DoLP, and AoP. Images in the white boxes show the magnification of local regions. White ellipses mark the differences of the reconstruction results the OLSM and SRM at defect pixels.
	}
	\label{fig8}
\end{figure}

The modulation parameters of the self-developed DoFP polarimeter are shown in Fig.~\ref{fig2}. The spectrum of the DoFP image captured by this polarimeter is shown in Fig.~\ref{fig3}(b), which is disturbed by the non-uniformity. The first three rows of Fig.~\ref{fig8} show the $s_0$, DoLP, and AoP reconstructed by the bicubic interpolation-based method, Newton's polynomial interpolation-based method, Planck-Taper window filtering-based method. The window function parameters used in the reconstruction were the same as the previous. The reconstruction results of these three methods contain strong non-uniformity artifacts, which indicates the interpolation-based and filtering-based methods fail in the presence of the non-uniformity. The fourth and fifth rows of Fig.~\ref{fig8} show the $s_0$, DoLP, and AoP reconstructed by the OLSM and SRM, respectively, with the non-uniform modulation parameters used in the reconstructions. The regularization parameters were chosen as "$\lambda_0=0.001$, $\lambda_1=0.0407$, $\lambda_2=0.0285$" according to the discussions given in Section~\ref{sec5}. Since the non-uniformity has been effectively characterized by the non-uniform modulation, the non-uniformity artifacts in the reconstruction results of these two methods are well mitigated. This proves the effectiveness of our methods for reconstructing the Stokes parameters in the presence of the non-uniformity. Comparing the reconstruction results of the OLSM and SRM in Fig.~\ref{fig7} and Fig.~\ref{fig8}, it can be seen that the performance of the OLSM and SRM in  these two experiments is basically consistent. In Fig.~\ref{fig8}, $s_0$ reconstructed by the OLSM is still blurred, and the reconstructed DoLP and AoP also show obviously serrated artifacts. While the $s_0$ reconstructed by the SRM shows more details, and the reconstructed DoLP and AoP show better visual effects. The reason for these phenomena is that when the non-uniformity has been considered in the reconstruction, the frequency responses of the two methods for the different frequency components of the Stokes parameters under the non-uniform modulation is basically the same as that under the ideal modulation. Notice that although the OLSM and SRM tackle the non-uniformity based on the same pixel model, the results are different at defect pixels. This phenomenon is marked by the white ellipses in Fig.~\ref{fig8}. It can be seen that the reconstruction results of the OLSM show more steep changes in flat background compared with that of the SRM. The performance of the defects pixels far deviates from the design performance. Since two adjacent defect pixels can easily cause the degeneration of the $4\times3$ matrix in Eq.(\ref{eq4}), the OLSM is more likely to fail at defect pixels. In contrast, the SRM is more robust to the defect pixels, this is due to the use of the global measurement information. However, empirically, we found that it is still difficult for the SRM to obtain satisfactory reconstruction results when the size of the defect area exceeds that of the biharmonic operator.

\section{Conclusion}\label{sec7}
We proposed the OLSM and SRM to reconstruct the Stokes parameters. The performance of the OLSM and SRM was investigated through Fourier analysis, numerical simulations, and experiments. The proposed methods can effectively mitigate the reconstruction errors and artifacts caused by the non-uniformity. The OLSM reconstructs the Stokes parameters under the local constant assumption. This method can be regraded as a generalization of the bilinear interpolation-based methods. Since the pseudo-inverse of the matrix in Eq.~\ref{eq4} can be pre-calculated, the OLSM consumes less computational costs, and is therefore more suitable for hardware implementation. The performance of the OLSM can be further improved if an optical low-passed filter is applied to restrict the bandwidths occupied by the frequency components of the Stokes parameters. The SRM reconstructs the Stokes parameters under the global smoothing assumption. This method has more flexible filtering performance. With a suitable choice of the regularization parameters, the SRM can reconstruct the Stokes parameters with good visual effects and low reconstruction errors. Some edge-preserved constraints can be applied to further improve the performance of the SRM.

\appendix
\section{Scaling and rotation transformation}\label{app:A}
Let $m_0'$, $m_1'$, and $m_2'$ be the original modulation parameters obtained from the calibration, the modulation parameters used in this paper are obtained by applying a scaling and rotation transformation to $m_0'$, $m_1'$, and $m_2'$, expressed as
\begin{equation}\label{eq19}
\begin{bmatrix}
	{{m_0}(x,y)} \\ 
	{{m_1}(x,y)} \\ 
	{{m_2}(x,y)} 
\end{bmatrix}
= \frac{1}{m}
\begin{bmatrix}
	1&0&0 \\ 
	0&{\cos (2\delta )}&{ - \sin (2\delta )} \\ 
	0&{\sin (2\delta )}&{\cos (2\delta )} 
\end{bmatrix}
\begin{bmatrix}
	{{m_0}'(x,y)} \\ 
	{{m_1}'(x,y)} \\ 
	{{m_2}'(x,y)} 
\end{bmatrix}.
\end{equation}
Here, $m$ is the mean of $m_0'$, and $\delta$ is the mean of the differences between the designed and the actual orientations of the linear polarization filters. Notice that the differences between the designed and the actual orientations should be wrapped between $-\pi/2$ and $\pi/2$. The purpose of this transformation is to normalize the modulation parameters and choose a relative orientation in the linear polarization filter array itself as the reference of 0$^{\circ}$. The changes caused by this transformation can be compensated by applying the inverse of this transformation to the reconstructed Stokes parameters.

\section{Explanation of numerical calibration}\label{app:D}
Numerical calibration methods are proposed to correct the non-uniform intensity responses in the DoFP image to ideal. A commonly used numerical calibration method is Gruev et. al.'s super-pixel calibration method, which can be expressed as
\begin{equation}
\begin{bmatrix}
\begin{smallmatrix}
{\hat i(2p,2q)} \\ 
{\hat i(2p,2q + 1)} \\ 
{\hat i(2p + 1,2q)} \\ 
{\hat i(2p + 1,2q + 1)} 
\end{smallmatrix}
\end{bmatrix}
= 
\begin{bmatrix}
\begin{smallmatrix}
1&1&0 \\ 
1&0&1 \\ 
1&-1&0 \\ 
1&0&-1 
\end{smallmatrix}
\end{bmatrix}
\underbrace{
	\begin{bmatrix}
	\begin{smallmatrix}
	{{m_0}(2p,2q)}&{{m_1}(2p,2q)}&{{m_2}(2p,2q)} \\ 
	{{m_0}(2p,2q + 1)}&{{m_1}(2p,2q + 1)}&{{m_2}(2p,2q + 1)} \\ 
	{{m_0}(2p + 1,2q)}&{{m_1}(2p + 1,2q)}&{{m_2}(2p + 1,2q)} \\ 
	{{m_0}(2p + 1,2q + 1)}&{{m_1}(2p + 1,2q + 1)}&{{m_2}(2p + 1,2q + 1)} 
	\end{smallmatrix}
	\end{bmatrix}
	^\dag
	\begin{bmatrix}
	\begin{smallmatrix}
	{i(2p,2q)} \\ 
	{i(2p,2q + 1)} \\ 
	{i(2p + 1,2q)} \\ 
	{i(2p + 1,2q + 1)} 
	\end{smallmatrix}
	\end{bmatrix}
}_{\text{Stokes parameters reconstruction}}
\end{equation}
Here, $\hat i$ represents the estimate of the ideal DoFP image, $p$ and $q$ are the coordinates which take half the ranges of $x$ and $y$, respectively. The super-pixel calibration method can be divided into two steps. The first step is also to reconstruct the Stokes parameters in the presence of the non-uniformity. The second step is to multiply the reconstructed results by a matrix defined by the ideal modulation parameters to generate $\hat i$. The Stokes parameters reconstruction in the super-pixel calibration method is similar to Eq.~\ref{eq4}. This indicates that the super-pixel calibration methods will definitely cause additional correction errors due to the spatial variations of the Stokes parameters.

Some studies suggest to first applying the super-pixel calibration methods to correct the non-uniformity, and then use the interpolation-based or filtering-based methods to reconstruct the Stokes parameters. However, we believe that this suggestion has potential problems. Firstly, the whole reconstruction process will be redundant, since the Stokes parameters are actually reconstructed twice. Secondly, with the advancement of the manufacturing technique, the actual modulation parameters of the DoFP polarimeter will be close to the ideal modulation parameters, it is foreseeable that the correction errors introduced by the super-pixel calibration method will exceed the non-uniformity errors mitigated by this method. In fact, the OLSM and SRM can also be used as the numerical calibration methods, achieved by substituting the ideal modulation parameters and the reconstructed Stokes parameters into Eq.~\ref{eq1} to generate $\hat i$. But based on the above concerns, we do not suggest to use the OLSM and SRM as the numerical calibration methods.

\section{Convolutional interpolation-based methods}\label{app:B}

\setcounter{table}{0}
\setcounter{figure}{0}

The Stokes parameters reconstructed by the convolutional interpolation-based methods can be expressed as
\begin{equation}\label{eq20}
\left\{
\begin{array}{l}
{{\hat s}_0} = i * \frac{1}{4}h \\
{{\hat s}_1} = (i \circ {m_1}) * \frac{1}{2}h \\
{{\hat s}_2} = (i \circ {m_2}) * \frac{1}{2}h
\end{array}
\!.\right.
\end{equation}
Here, $\circ$ represents the Hadamard product, $m_1$ and $m_2$ take the ideal values given in Eq.~(\ref{eq3}), $h$ represents the interpolation kernel. The interpolation kernels of the nearest-neighbor, bilinear, bicubic, and natural bicubic spline interpolation algorithms are given in Table.~\ref{tab3}. The natural bicubic spline interpolation algorithm is a piece-wise interpolation algorithm, therefore the interpolation kernel can have different sizes. The $7\times7$ and $11\times11$ sized interpolation kernels are given as example.

\begin{table}[]
	\centering
	\caption{Interpolation kernels.}
	\resizebox{\textwidth}{!}{
	\begin{tabular}{ll}
		\hline
		Interpolation algorithms	&   Interpolation kernel $h=ww^T$\\ \hline
		Nearest-neighbor	&  $w = {\left[ {\begin{array}{cc}1&1\end{array}} \right]^T}$ \\
		Bilinear	& $w = {\left[ {\begin{array}{ccc}\tfrac{1}{2} & 1 & \tfrac{1}{2}\end{array}} \right]^T}$ \\
		Bicubic	& $w = {\left[ {\begin{array}{ccccccc}{ - \tfrac{1}{{16}}}&0&{\tfrac{9}{{16}}}&1&{\tfrac{9}{{16}}}&0&{ - \tfrac{1}{{16}}}\end{array}} \right]^T}$ \\
		natural bicubic spline 7$\times$7	& $w = {\left[ {\begin{array}{ccccccc}{ - \tfrac{3}{{40}}}&0&{\tfrac{{23}}{{40}}}&1&{\tfrac{{23}}{{40}}}&0&{ - \tfrac{3}{{40}}}\end{array}} \right]^T}$ \\
		natural bicubic spline 11$\times$11	& $w = {\left[ {\begin{array}{ccccccccccc}{\tfrac{3}{{152}}}&0&{ - \tfrac{9}{{76}}}&0&{\tfrac{{91}}{{152}}}&1&{\tfrac{{91}}{{152}}}&0&{ - \tfrac{9}{{76}}}&0&{\tfrac{3}{{152}}}\end{array}} \right]^T}$ \\ \hline
	\end{tabular}}
	\label{tab3}
\end{table}

\begin{figure}[hbt!]
	\centering\includegraphics[width=12cm]{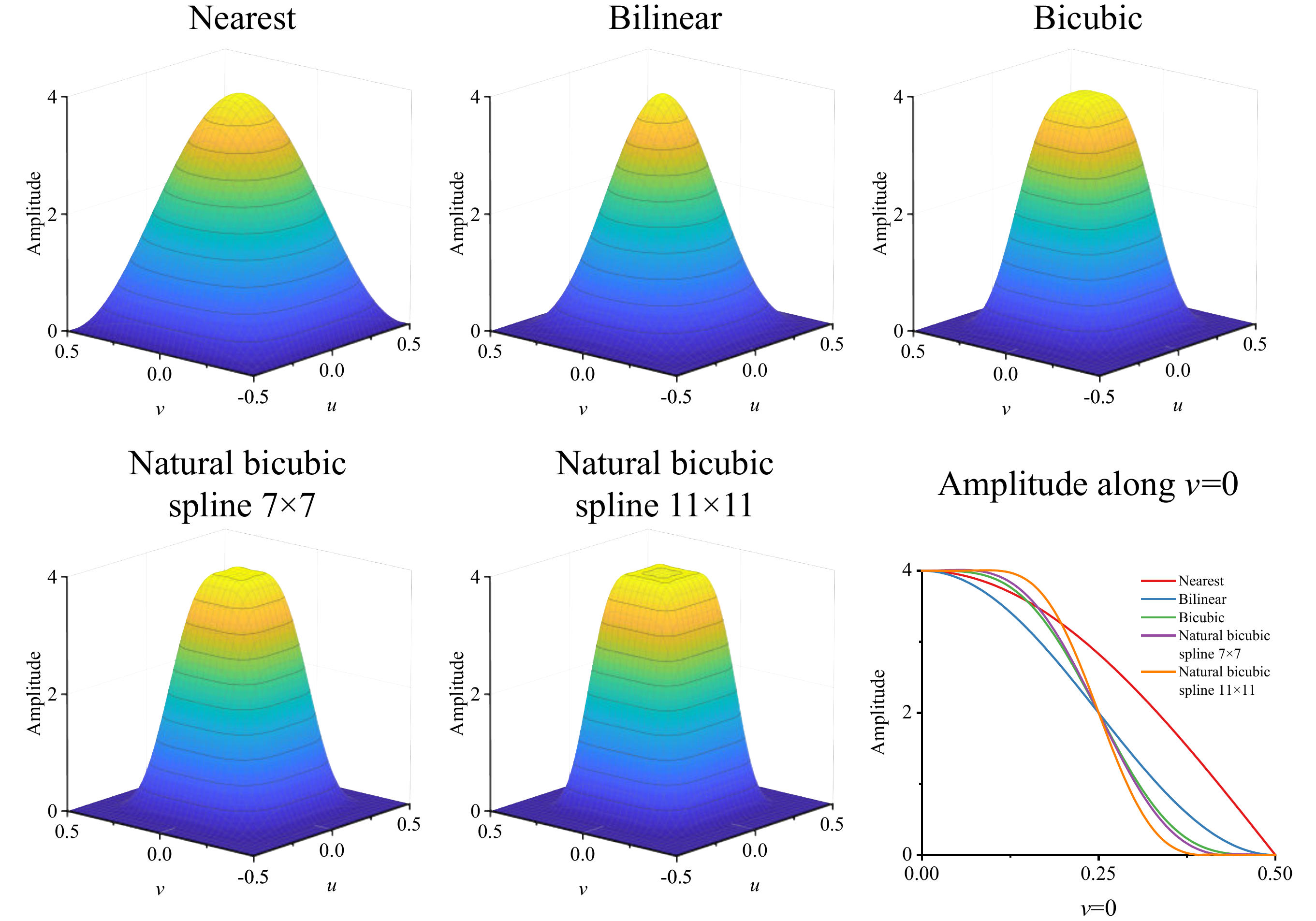}
	\caption{DFTs of the interpolation kernels of the nearest-neighbor, bilinear, bicubic, and natural bicubic spline interpolation algorithms and their amplitudes along $v=0$.}
	\label{fig9}
\end{figure}

Performing the DFT of Eq.~(\ref{eq20}), after some simplifications, we have
\begin{equation}\label{eq21}
\left\{
\begin{array}{l}
{{\hat S}_0}(u,v) = I(u,v) \cdot \frac{1}{4}H(u,v) \\
\tfrac{1}{2}[{{\hat S}_1}(u + \tfrac{1}{2},v) + {{\hat S}_2}(u + \tfrac{1}{2},v)] = I(u,v) \cdot \frac{1}{4}H(u + \tfrac{1}{2},v) \\
\tfrac{1}{2}[{{\hat S}_1}(u,v + \tfrac{1}{2}) - {{\hat S}_2}(u,v + \tfrac{1}{2})] = I(u,v) \cdot \frac{1}{4}H(u,v + \tfrac{1}{2})
\end{array}
\!.\right.
\end{equation}
The DFTs of the interpolation kernels of the nearest-neighbor, bilinear, bicubic, and natural bicubic spline interpolation algorithms and their amplitudes along $v=0$ are plotted in Fig.~\ref{fig9}. It can be found that the interpolation kernels of the bilinear, bicubic, and natural bicubic spline interpolation algorithms are the approximations of $\text{sinc}(x/2)\text{sinc}(y/2)$ in the spatial domain, and the DFTs are the approximations of $4\text{rect}(2u)\text{rect}(2v)$ in the frequency domain.

\section{Planck-Taper window function}\label{app:C}
\setcounter{figure}{0}

\begin{figure}[hbt!]
	\centering\includegraphics[width=\textwidth]{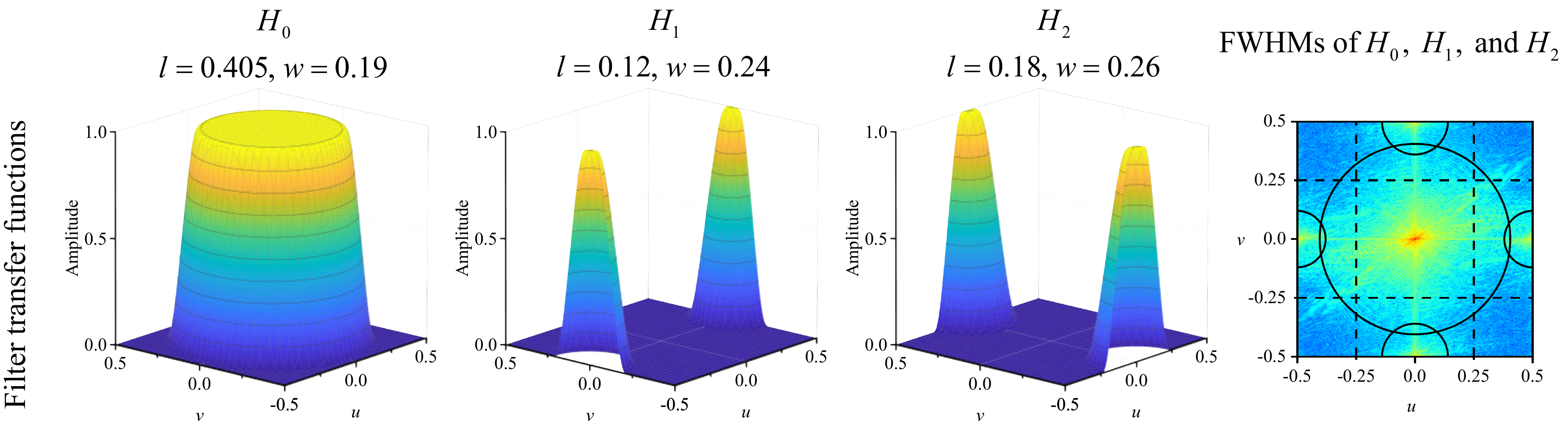}
	\caption{Filter transfer functions constructed by the Planck-Taper window function.}
	\label{fig10}
\end{figure}

The Planck-Taper window function is defined as~\cite{RN1051}
\begin{equation}\label{eq22}
H(r) = \left\{ {\begin{array}{cc}
	{1,}&{r \leq l - w/2} \\ 
	{\frac{1}{{1 + \exp \left( { - \frac{w}{{r - l + w/2}} - \frac{w}{{r - l - w/2}}} \right)}},}&{l - w/2 < r < l + w/2} \\ 
	{0,}&{r \geq l + w/2} 
	\end{array}}\! .\right.
\end{equation}
Here, $r=\sqrt{u^2+v^2}$, $l$ is the radius of the FWHM, $w$ is the falloff range. The filter transfer functions constructed by the Planck-Taper window function used in Section \ref{sec6} are illustrated in Fig.~\ref{fig10}.

\section*{Acknowledgements}
This work was supported by the National Natural Science Foundation of China (Grant Nos. 11627803, 11872354, 11872355) and the Strategic Priority Research Program of the Chinese Academy of Sciences (XDB22040502).

\bibliography{sample}

\begin{thebibliography}{10}
\expandafter\ifx\csname url\endcsname\relax
  \def\url#1{\texttt{#1}}\fi
\expandafter\ifx\csname urlprefix\endcsname\relax\def\urlprefix{URL }\fi
\expandafter\ifx\csname href\endcsname\relax
  \def\href#1#2{#2} \def\path#1{#1}\fi

\bibitem{RN1742}
J.~S. Tyo, D.~L. Goldstein, D.~B. Chenault, J.~A. Shaw, Review of passive
  imaging polarimetry for remote sensing applications, Applied optics 45~(22)
  (2006) 5453--5469.

\bibitem{RN1741}
S.~Alali, I.~A. Vitkin, Polarized light imaging in biomedicine: emerging
  mueller matrix methodologies for bulk tissue assessment, Journal of
  biomedical optics 20~(6) (2015) 061104.

\bibitem{RN1745}
Q.-H. Phan, Y.-L. Lo, Stokes-mueller matrix polarimetry system for glucose
  sensing, Optics and Lasers in Engineering 92 (2017) 120--128.

\bibitem{RN93}
Z.~G. Zhang, F.~L. Dong, K.~M. Qian, Q.~C. Zhang, W.~G. Chu, Y.~T. Zhang,
  X.~Ma, X.~P. Wu, Real-time phase measurement of optical vortices based on
  pixelated micropolarizer array, Optics Express 23~(16) (2015) 20521--20528.

\bibitem{RN1546}
Y.~Zhang, F.~Dong, K.~Qian, Q.~Zhang, W.~Chu, X.~Ma, X.~Wu, Study on evolving
  phases of accelerating generalized polygon beams, Optics Express 24~(5)
  (2016) 5300--5310.

\bibitem{RN1545}
Y.~Zhang, Q.~Zhang, X.~Ma, Z.~Jiang, T.~Xu, S.~Wu, X.~Wu, Measurement of
  airy-vortex beam topological charges based on a pixelated micropolarizer
  array, Applied Optics 55~(32) (2016) 9299--9304.

\bibitem{RN1748}
Q.~Kemao, M.~Hong, W.~Xiaoping, Real-time polarization phase shifting technique
  for dynamic deformation measurement, Optics and lasers in engineering 31~(4)
  (1999) 289--295.

\bibitem{RN1747}
H.-K. Teng, K.-C. Lang, Polarization shifting interferometric profilometer,
  Optics and Lasers in Engineering 46~(3) (2008) 203--210.

\bibitem{RN1749}
P.~Yan, K.~Wang, P.~Cui, J.~Gao, J.~Ma, Q.~Zhang, Single-exposure polarization
  phase-shifting interferometer using an azo-polymer orientation array, Optics
  and Lasers in Engineering 73 (2015) 75--79.

\bibitem{RN1736}
C.~K. Harnett, H.~G. Craighead, Liquid-crystal micropolarizer array for
  polarization-difference imaging, Applied optics 41~(7) (2002) 1291--1296.

\bibitem{RN1737}
E.~Compain, B.~Drevillon, Broadband division-of-amplitude polarimeter based on
  uncoated prisms, Applied optics 37~(25) (1998) 5938--5944.

\bibitem{RN1746}
W.-C. Liu, Y.-L. Lo, Q.-H. Phan, Circular birefringence/dichroism measurement
  of optical scattering samples using amplitude-modulation polarimetry, Optics
  and Lasers in Engineering 102 (2018) 45--51.

\bibitem{RN1735}
W.~Zhang, J.~Liang, L.~Ren, H.~Ju, E.~Qu, Z.~Bai, Y.~Tang, Z.~Wu, Real-time
  image haze removal using an aperture-division polarimetric camera, Applied
  optics 56~(4) (2017) 942--947.

\bibitem{RN833}
G.~P. Nordin, J.~T. Meier, P.~C. Deguzman, M.~W. Jones, Micropolarizer array
  for infrared imaging polarimetry, Journal of the Optical Society of America
  a-Optics Image Science and Vision 16~(5) (1999) 1168--1174.

\bibitem{RN1725}
Z.~Zhang, F.~Dong, T.~Cheng, K.~Qian, K.~Qiu, Q.~Zhang, W.~Chu, X.~Wu, Electron
  beam lithographic pixelated micropolarizer array for real-time phase
  measurement, Chinese Physics Letters 31~(11) (2014) 114208--114208.

\bibitem{RN1726}
Z.~Zhang, F.~Dong, T.~Cheng, K.~Qiu, Q.~Zhang, W.~Chu, X.~Wu, Nano-fabricated
  pixelated micropolarizer array for visible imaging polarimetry, Review of
  scientific instruments 85~(10) (2014) 105002.

\bibitem{RN1724}
B.~M. Ratliff, C.~F. LaCasse, J.~S. Tyo, Interpolation strategies for reducing
  ifov artifacts in microgrid polarimeter imagery, Optics express 17~(11)
  (2009) 9112--9125.

\bibitem{RN836}
S.~K. Gao, V.~Gruev, Bilinear and bicubic interpolation methods for division of
  focal plane polarimeters, Optics Express 19~(27) (2011) 26161--26173.

\bibitem{RN1652}
N.~Li, Y.~Zhao, Q.~Pan, S.~G. Kong, Demosaicking dofp images using newton's
  polynomial interpolation and polarization difference model, Optics Express
  27~(2) (2019) 1376--1391.

\bibitem{RN1593}
S.~Mihoubi, P.-J. Lapray, L.~Bigué, Survey of demosaicking methods for
  polarization filter array images, Sensors 18~(11) (2018) 3688.

\bibitem{RN1047}
J.~S. Tyo, C.~F. LaCasse, B.~M. Ratliff, Total elimination of sampling errors
  in polarization imagery obtained with integrated microgrid polarimeters,
  Optics Letters 34~(20) (2009) 3187--3189.

\bibitem{RN1665}
D.~A. LeMaster, K.~Hirakawa, Improved microgrid arrangement for integrated
  imaging polarimeters, Optics Letters 39~(7) (2014) 1811--1814.

\bibitem{RN1051}
A.~S. Alenin, I.~J. Vaughn, J.~Scott~Tyo, Optimal bandwidth micropolarizer
  arrays, Optics Letters 42~(3) (2017) 458--461.

\bibitem{RN1502}
D.~L. Bowers, J.~K. Boger, L.~D. Wellems, S.~E. Ortega, M.~P. Fetrow, J.~E.
  Hubbs, W.~T. Black, B.~M. Ratliff, J.~S. Tyo, Unpolarized calibration and
  nonuniformity correction for long-wave infrared microgrid imaging
  polarimeters, Optical Engineering 47~(4).

\bibitem{RN834}
S.~B. Powell, V.~Gruev, Calibration methods for division-of-focal-plane
  polarimeters, Optics Express 21~(18) (2013) 21039--21055.

\bibitem{RN1720}
N.~A. Hagen, S.~Shibata, Y.~Otani, Calibration and performance assessment of
  microgrid polarization cameras, Optical Engineering 58~(8) (2019) 1--9, 9.

\bibitem{RN1727}
B.~D. Lucas, T.~Kanade, An iterative image registration technique with an
  application to stereo vision, in: In IJCAI81, 1981, pp. 674--679.

\bibitem{RN1728}
B.~K. Horn, B.~G. Schunck, Determining optical flow, Artificial intelligence
  17~(1-3) (1981) 185--203.

\bibitem{david2008unpolarized}
D.~L. Bowers, J.~K. Boger, D.~Wellems, S.~Ortega, M.~P. Fetrow, J.~E. Hubbs,
  W.~T. Black, B.~M. Ratliff, J.~S. Tyo, {Unpolarized calibration and
  nonuniformity correction for long-wave infrared microgrid imaging
  polarimeters}, Optical Engineering 47~(4) (2008) 1 -- 9.

\bibitem{RN1643}
B.~Feng, Z.~Shi, H.~Liu, L.~Liu, Y.~Zhao, J.~Zhang, Polarized-pixel performance
  model for dofp polarimeter, Journal of Optics 20~(6).

\bibitem{RN1637}
X.~Ma, F.~Dong, Z.~Zhang, Y.~Su, T.~Xu, Z.~Jiang, S.~Wu, Q.~Zhang, W.~Chu,
  X.~Wu, Pixelated-polarization-camera-based polarimetry system for wide
  real-time optical rotation measurement, Sensors and Actuators B: Chemical 283
  (2019) 857--864.

\bibitem{boyd2004convex}
S.~Boyd, L.~Vandenberghe, Convex Optimization, Cambridge University Press,
  2004.

\bibitem{srm}
{Zhaoxiang Jiang}, {Smoothing regularization method for reconstructing Stokes
  parameters from division-of-focal-plane modulation},
  \url{https://github.com/zxcn/SRM} (2020).

\end{thebibliography}
\end{document}